\documentclass[11pt, a4paper]{article}

\usepackage{jcappub}
\usepackage{graphicx}
\usepackage{slashed}
\usepackage{booktabs}
\usepackage{multirow}
\usepackage{color}
\usepackage{textpos}

\newcommand{\kpc}{\ \text{kpc}}
\newcommand{\GeV}{\ \text{GeV}}
\newcommand{\s}{\ \text{s}}
\newcommand{\cm}{\ \text{cm}}

\newcommand{\fex}{e.g.~}
\newcommand{\ie}{\textit{i.e.}~}
\newcommand{\cf}{\textit{cf.}~}

\title{Fermi LAT Search for Internal Bremsstrahlung Signatures from Dark
Matter Annihilation}

\author[a]{Torsten Bringmann}
\author[b]{Xiaoyuan Huang}
\author[c]{Alejandro Ibarra}
\author[c]{Stefan Vogl}
\author[d]{Christoph Weniger}

\affiliation[a]{II. Institute for Theoretical Physics, University of Hamburg,
Luruper Chaussee 149, DE-22761 Hamburg, Germany}
\affiliation[b]{National Astronomical Observatories, Chinese Academy of
Sciences, Beijing, 100012, China}
\affiliation[c]{Physik-Department T30d, Technische Universit\"at M\"unchen,
James-Franck-Stra\ss{}e, 85748 Garching, Germany}
\affiliation[d]{Max-Planck-Institut f\"ur Physik, F\"ohringer Ring 6, 80805
Munich, Germany}

\emailAdd{torsten.bringmann@desy.de}
\emailAdd{x\_huang@bao.ac.cn}
\emailAdd{ibarra@tum.de}
\emailAdd{stefan.vogl@tum.de}
\emailAdd{weniger@mppmu.mpg.de}

\abstract{A commonly encountered obstacle in indirect searches for galactic
dark matter is how to disentangle possible signals from astrophysical
backgrounds. Given that such signals are most likely subdominant, the search
for pronounced spectral features plays a key role for indirect detection
experiments; monochromatic gamma-ray lines or similar features related to
internal bremsstrahlung, in particular, provide smoking gun signatures. We
perform a dedicated search for the latter in the data taken by the Fermi
gamma-ray space telescope during its first 43 months. To this end, we use a
new adaptive procedure to select optimal target regions that takes into
account both standard and contracted dark matter profiles. The behaviour of
our statistical method is tested by a subsampling analysis of the full sky
data and found to reproduce the theoretical expectations very well.  The
limits on the dark matter annihilation cross-section that we derive are
stronger than what can be obtained from the observation of dwarf galaxies and,
at least for the model  considered here, collider searches.  While these
limits are still not quite strong enough to probe annihilation rates expected
for thermally produced dark matter, future prospects to do so are very good.
In fact, we already find a weak indication, with a significance of $3.1\sigma$
($4.3\sigma$) when (not) taking into account the look-elsewhere effect,  for
an internal bremsstrahlung-like signal that would correspond to a dark matter
mass of $\sim$150\,GeV; the same signal is also well  fitted by a gamma-ray
line at around 130 GeV.  Although this would be a fascinating possibility, we
caution that a much more dedicated analysis and additional data will be
necessary to rule out or confirm this option.}

\begin{document}
\begin{textblock}{3}(9.3,0)
  \noindent
  TUM-HEP 828/12\\
  MPP-2012-54
\end{textblock}

\maketitle

\section{Introduction}
While there is little doubt about the very existence of a cold dark matter
(DM) component in the present composition of the universe, contributing a
substantial  fraction of $\Omega_\chi = 0.229 \pm 0.015$ to its total energy
budget \cite{Komatsu:2010fb}, the DM nature  remains unknown even 80 years
after Zwicky first postulated a 'missing mass' in the Coma cluster
\cite{Zwicky:1933gu}. At the moment, the leading hypothesis for a solution to
this puzzle are  thermally produced, weakly interacting massive particles
(WIMPs) as expected to appear in many extensions to the standard model of
particle physics (see, e.g., Refs.~\cite{Jungman:1995df, Bergstrom:2000pn,
Bertone:2004pz} for reviews or Ref.~\cite{Bergstrom:2012np} for a recent
pedagogical introduction) -- though one should keep in mind that the absence
of any clear signs for new physics at the CERN LHC may render this theoretical
prejudice less likely already in a few years from now \cite{Bertone:2010at}.
If DM consists of WIMPs, it can be searched for both at \emph{colliders} (with
missing transverse energy as the main signature), in \emph{direct} detection
experiments looking for the scattering of WIMPs with the nuclei of terrestrial
detectors, or \emph{indirectly} through the observation of WIMP annihilation
products in cosmic rays; the obvious advantage of indirect searches (see
\cite{Cirelli:2012tf} for a recent status review) being that they allow to
probe the nature of DM  not only locally but on cosmological (or at least
galactic) scales.

\emph{Gamma rays} are for several reasons a particularly suitable detection
channel for DM annihilation, not the least because they propagate essentially
unhindered through the galaxy and thus directly point back to the sources.
Here, we will focus on another important aspect, namely the possibility of
\emph{sharp spectral features} in the annihilation spectra that would allow a
rather straightforward discrimination from astrophysical backgrounds. Indeed,
it has early been pointed out~\cite{Bergstrom:1988fp} that the direct
annihilation into photons would lead to the smoking gun signature of a
gamma-ray line~\cite{Bergstrom:1997fh, Ullio:1997ke, Bern:1997ng,
Bergstrom:2004nr,Gustafsson:2007pc,
Mambrini:2009ad,Jackson:2009kg,Bertone:2009cb, Bertone:2010fn}. 
Equally pronounced spectral
features near photon energies close to the kinematical endpoint at
$E_\gamma=m_\chi$ arise due to \emph{internal bremsstrahlung}, i.e.~in the
presence of three-body final states containing a
photon~\cite{Beacom:2004pe,Birkedal:2005ep,Bergstrom:2004cy,Bringmann:2007nk,
Barger:2009xe, Barger:2011jg}, albeit at much higher rates because they are
not loop-suppressed.  Including such features in the analysis, rather than
following the common approach of looking for featureless spectral templates,
can significantly increase the sensitivity to DM
signals~\cite{Bringmann:2008kj,Bringmann:2011ye} and might even be used to
rather efficiently discriminate between different DM models (see
e.g.~Ref.~\cite{Perelstein:2010at}).

The Large Area Telescope (LAT)~\cite{Atwood:2009ez}, the main instrument on
board the Fermi Gamma-ray space telescope, has an unprecedented sensitivity to
gamma rays from 30\,MeV to above 300\,GeV. This, together with its large
field-of-view, makes it ideally suited for DM searches~\cite{Baltz:2008wd,
Abdo:2010ex, Ackermann:2011wa, Ackermann:2010rg, Abdo:2010dk, Scott:2009jn,
Abazajian:2010sq, Abdo:2010nc, Vertongen:2011mu, Huang:2011xr,
GeringerSameth:2011iw}; the non-observation of gamma-ray signals from dwarf
galaxies, e.g., places the currently most stringent bounds on the annihilation
rate of WIMPs with masses below around 700\,GeV \cite{Abdo:2010ex,
Ackermann:2011wa, GeringerSameth:2011iw} (for higher masses,
H.E.S.S.~observations of the galactic center lead to stronger limits
\cite{Abramowski:2011hc}). So far, only line-signals have been searched for in
the LAT \cite{Abdo:2010nc, Vertongen:2011mu} or EGRET \cite{Pullen:2006sy}
data; the main purpose of this article is to extend these searches to other
spectral endpoint features which, as outlined above, are arguably more
relevant in most WIMP models. In order to look for such templates, we adopt a
strategy that is very close in spirit to traditional line searches, as
recently shown to be very promising \cite{Bringmann:2011ye}, and use a new
adaptive method to identify optimized target regions around the galactic
center. As we will show, this leads to very competitive constraints on
possible annihilation signals. 

For Majorana (but also scalar) WIMPs, the pair of annihilating DM particles
approximately forms a $J=0$ state for the small relative velocities
$v\sim10^{-3}$ expected in the galactic halo. In this commonly encountered
situation, the annihilation rate into fermionic two-body final states $\bar f
f$ is (usually quite strongly) suppressed by $m_f^2/m_\chi^2$ and the next
order result---with an additional photon in the final state---can be greatly
enhanced by a factor of $\sim(\alpha_{\rm em}/\pi)m_\chi^2/m_f^2$; since this
only happens for $t$- and $u$-channel annihilation mediated by the exchange of
charged virtual  particles, this process is also referred to as
\emph{virtual} internal bremsstrahlung (VIB, see Ref.~\cite{Bringmann:2007nk}
for an extensive discussion).  Apart from the expected large rates, VIB also
results in very pronounced spectral features close to the kinematic endpoint
which resemble a slightly distorted line.  Taken together, these two aspects
make VIB in some sense the most promising DM signature to look for and
motivate our choice of mainly focussing  on related spectral distortions in
the astrophysical background. 

In order not to obscure our analysis by the potentially many  parameters
entering into the DM model, we will focus on a simple toy model that
corresponds to the minimal extension to the standard model where we can expect
strong VIB signals. While this model has been considered before and in its own
right---in particular in connection with electroweak bremsstrahlung
corrections to the annihilation rate \cite{Kachelriess:2009zy, Bell:2011eu,
Ciafaloni:2011sa, Bell:2011if, Garny:2011cj, Ciafaloni:2011gv, Garny:2011ii},
but also in other contexts \cite{Ma:2000cc, Cao:2009yy}---let us stress that
our analysis can straight-forwardly be applied to any other  model with
similar  gamma-ray spectra from DM annihilation; in particular, the main
features of this model are the same, for our purpose, as for neutralino DM in
some phenomenologically very relevant  regions of the supersymmetric parameter
space.

The structure of this article is as follows. In Section \ref{sec:model}, we
introduce our toy model and comment on how it relates to the more commonly
considered case of supersymmetric neutralino DM.  We describe our method to
search for pronounced spectral templates in the LAT data in Section
\ref{sec:method}  and also present our results there. In Section
\ref{sec:constraints}, we compare our new limits to existing limits from dwarf
galaxies, expectations for thermally produced DM, collider constraints and
limits from cosmic-ray anti-protons. We present our conclusions in Section
\ref{sec:conc}.  Finally, we provide some additional technical information
about our method of selecting a target region optimized for the search of
DM-related spectral features (Appendix \ref{apx:targetRegions}) and how a
subsampling analysis of the full sky data can be used as a further
test to confirm the reliability of our statistical method  (Appendix
\ref{apx:details}).

\section{Particle physics scenario}
\label{sec:model}
\subsection{Toy model with large virtual internal bremsstrahlung}
We will assume that the DM of the Universe is constituted by Majorana fermions
$\chi$, singlets under the Standard Model gauge group, which couple to the
Standard Model via a Yukawa interaction with a scalar $\eta$ that is  not much heavier
than the DM particle.  The Lagrangian of the model reads:
\begin{align}
  {\cal L}={\cal L}_{\rm SM}+{\cal L}_{\chi}+{\cal L}_\eta+ {\cal L}_{\rm int}
  \;.
\end{align}
Here, ${\cal L}_{\rm SM}$ is the Standard Model Lagrangian.  ${\cal L}_\chi$
and ${\cal L}_\eta$ are the parts of the Lagrangian involving only the
Majorana fermion $\chi$ and the scalar particle $\eta$, respectively, and  are
given by
\begin{align}
  \begin{split}
    {\cal L}_\chi&=\frac12 \bar \chi^c i\slashed {\partial} \chi
    -\frac{1}{2}m_\chi \bar \chi^c\chi\;, \\ {\cal L}_\eta&=(D_\mu
    \eta)^\dagger  (D^\mu \eta)-m_\eta^2 \eta^\dagger\eta \;,
  \end{split}
\end{align}
where $D_\mu$ denotes the covariant derivative. Lastly, ${\cal L}_{\rm int}$
denotes the interaction terms of the new particles with Standard Model fields.

We will consider in this paper three toy models where the DM particle only
couples to the right-handed muons, tau leptons or bottom quarks, respectively,
via a Yukawa interaction with the scalar $\eta$. We assume the latter to be an
$SU(2)_L$ singlet in order to avoid  constraints from electroweak precision
measurements. The gauge quantum numbers of the intermediate scalar $\eta$ are
$(\mathbf{1},\mathbf{1})_1$  for couplings with the muon or the tau
(\textit{i.e.}~$\eta$ is a $SU(3)_c$ and $SU(2)_L$ singlet with hypercharge
$Y=1$), and $(\mathbf{\bar3},\mathbf{1})_{1/3}$ for couplings with the bottom
quark.  Furthermore, to guarantee a coupling to just one generation of
fermions we assign $\eta$ a muon number $L_\mu=-1$, a tau number $L_\tau=-1$
or a beauty number $B=-1$, respectively. Then, the relevant part of the
interaction Lagrangian reads
\begin{align}
  {\cal L}_{\rm int} &= -y \bar \chi \Psi_R \eta+{\rm h.c.} \;,
\label{eq:singlet-eR}
\end{align}
with $\Psi=\mu,\,\tau,\, b$. Note that  in principle additional
couplings of the form $H^\dagger H\eta^\dagger\eta$ and $(\eta^\dagger\eta)^2$
are allowed (where $H$ denotes the Higgs doublet). We will neglect them throughout
this work since they do not directly influence the gamma-ray signature we are
interested in.

In these scenarios, DM particles can annihilate into two fermions with a
velocity-weighted annihilation cross-section which can be decomposed into an
$s$-wave and a $p$-wave contribution.  The $s$-wave contribution reads in
lowest order of the relative center-of-mass velocity $v$~ \cite{Ellis:1998kh,
Nihei:2002sc}
\begin{align}
  (\sigma v)_\text{2-body}^\text{$s$-wave}= \frac{y^4 N_c}{32\pi m_\chi^2}
  \frac{m_f^2}{m_\chi^2}\frac{1}{(1+\mu)^2}\;,
  \label{eqn:sv2s}
\end{align}
where $\mu\equiv(m_\eta/m_\chi)^2$ parametrizes the mass splitting between the
DM particle $\chi$ and the $t$-channel mediator $\eta$, and the color factor
$N_c$ is one for muons and taus and three for bottom quarks. The
$p$-wave contribution at lowest order in $v$ is~\cite{Cao:2009yy}
\begin{align}
  (\sigma v)_\text{2-body}^\text{$p$-wave} = v^2 \frac{y^4 N_c}{48\pi
  m_\chi^2}
  \frac{1+\mu^2}{(1+\mu)^4}\;.
  \label{eqn:sv2p}
\end{align}

It is important to note that the $s$-wave contribution to the
velocity-weighted annihilation cross-section of Majorana fermions is helicity
suppressed, by the mass squared of the daughter fermion, whereas  the $p$-wave
contribution is suppressed by the velocity squared of the galactic DM
particles today, typically $v\sim 10^{-3}$. Therefore, the $2\rightarrow 2$
annihilation cross-section is fairly small, and higher order corrections could
play an important role.

\begin{figure}[t]
  \begin{center}
    \includegraphics[width=1.0\linewidth]{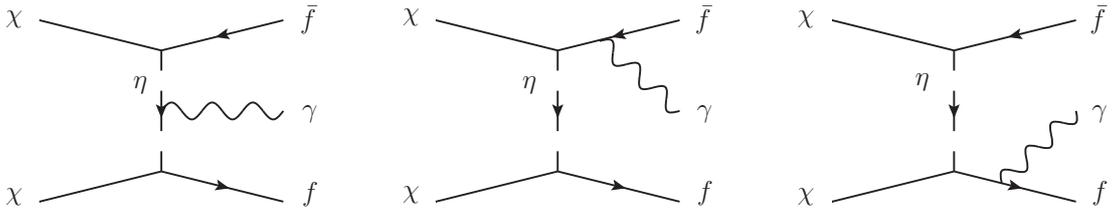}
  \end{center}
  \caption{Feynman diagrams of the processes that contribute in leading order
  to the three-body annihilation cross-section and produce internal
  bremsstrahlung. The first diagram very roughly corresponds to VIB, the
  second and third to FSR (but note that these contributions can be properly
  defined and separated in a gauge-invariant way \cite{Bringmann:2007nk}).}
  \label{fig:diagrams}
\end{figure}

Indeed, it was shown in Refs.~\cite{Bergstrom:1989jr, Flores:1989ru} that the
associated emission of a vector boson lifts the helicity suppression in the
$s$-wave contribution to the annihilation cross-section. This process was
later dubbed \emph{virtual internal bremsstrahlung} (VIB), which together with
photons from final-state radiation (FSR) constitute the full internal
bremsstrahlung (IB) spectrum \cite{Bringmann:2007nk}; the corresponding
Feynman diagrams are shown in Fig.~\ref{fig:diagrams}. The explicit expression
for the annihilation cross-section into two \emph{massless} fermions and one
VIB photon is~\cite{Bergstrom:2008gr, Bell:2011if}:
\begin{align}
  (\sigma v)_\text{3-body}
  \simeq  \frac{\alpha_\text{em} y^4 N_c Q_f^2}{64\pi^2m_\chi^2}
  & \left\{ 
  (\mu+1) \left[ \frac{\pi^2}{6}-\ln^2\left( \frac{\mu+1}{2\mu} \right)
  -2\text{Li}_2\left( \frac{\mu+1}{2\mu} \right)\right]  \right. \nonumber \\
   &+ \left. \frac{4\mu+3}{\mu+1}+\frac{4\mu^2-3\mu-1}{2\mu}\ln\left(
  \frac{\mu-1}{\mu+1} \right)
  \right\}\;,
  \label{eqn:sv3}
\end{align}
which is usually non-negligible and can in certain instances, in particular
for small values of $\mu$, be considerably larger than the $2\rightarrow 2$
annihilation cross-sections Eqns.~(\ref{eqn:sv2s},\ref{eqn:sv2p}); $Q_f$
denotes the electric charge of $f$ and $\eta$ in units of $|e|$. We emphasize
that throughout this paper ``3-body process'' refers to the VIB process only,
whereas ``2-body process'' refers to the helicity-suppressed tree-level
process $\chi\chi\to f \bar{f}$ plus the FSR photons. When explicitly referring 
to $(\sigma v)_\text{2-body}$ in the following, we will therefore multiply
 Eqs.~(\ref{eqn:sv2s},\ref{eqn:sv2p})  by a factor of 
$\left(1+\int dx\,dN^{\rm FSR}/dx\right)$, 
where \cite{Birkedal:2005ep}
\begin{equation}
\frac{dN^{\rm FSR}}{dx}=\frac{\alpha_{\rm em}
Q_f^2}{\pi}\frac{1+(1-x)^2}{x}\log
\left(\frac{4m_\chi^2(1-x)}{m_f^2}\right)\,.
\end{equation}

Furthermore, the energy spectrum of gamma rays produced in the $2\rightarrow
3$ process has a very peculiar shape that allows for an efficient search for
gamma rays from internal bremsstrahlung. Namely, the differential three-body
cross-section, as function of the VIB photon energy $x\equiv E/m_\chi$, is
given by~\cite{Bringmann:2007nk}
\begin{align}
  v\frac{d\sigma}{dx}
  \simeq  \frac{\alpha_\text{em} y^4 N_c Q_f^2}{32\pi^2m_\chi^2}
  (1-x)&\left\{\frac{2x}{(\mu+1)(\mu+1-2x)}-\frac{x}{(\mu+1-x)^2} \right. \nonumber \\
  & \left. -\frac{(\mu+1)(\mu+1-2x)}{2(\mu+1-x)^3}\ln
  \left( \frac{\mu+1}{\mu+1-2x} \right)
  \right\}\;.
  \label{eqn:sv3spec}
\end{align}

\begin{figure}[t]
  \centering
  \includegraphics[width=0.7\linewidth]{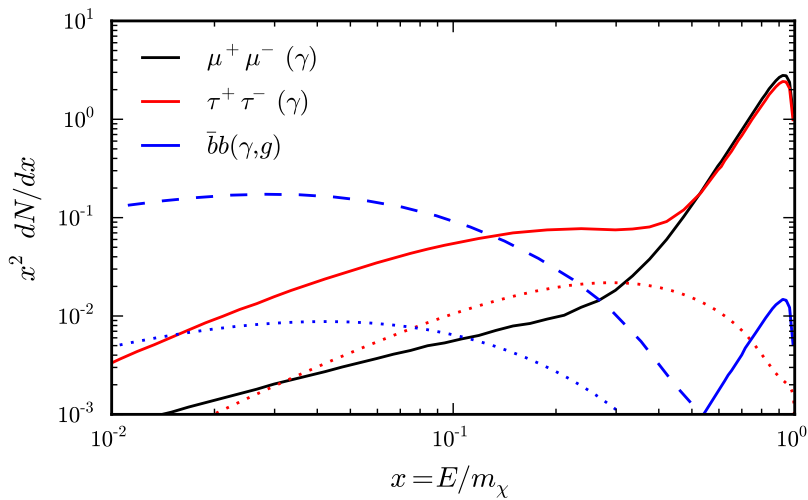}
  \vspace{-0.3cm}
  \caption{Gamma-ray spectrum ($N$ denotes the number of photons produced
  \emph{per annihilation}) as predicted by our toy model for different 
   final-state fermions, assuming $m_\chi=200\GeV$ and a mass-splitting of
  $\mu=1.1$. \emph{Solid lines} show the full contribution from three-body final
  states, including the VIB photons close to $x=1$; \emph{dotted lines} show
  contributions from the helicity-suppressed two-body final states including
  FSR (in case of muons, the latter is strongly suppressed and not visible on
  the plotted scales). Branching ratios are calculated according to
  Eqns.~\eqref{eqn:sv2s} and \eqref{eqn:sv3}. In case of bottom-quarks, we
  also include contributions from gluon VIB, $\chi\chi\to\bar{b}bg$, following
  Ref.~\cite{Asano:2011ik, Garny:2011ii} (\emph{dashed line}). Note that we
  convolve the spectra shown here with the Fermi LAT energy dispersion as
  derived from the instrument response functions (about $\Delta E\sim10\%$ at
  around 100~GeV~\cite{Fermi:performancePASS7}) before any fits to the data
  are performed.}
  \label{fig:spectra}
\end{figure}

To illustrate the peculiar features of VIB, the \emph{energy spectrum of gamma
rays} that is produced per annihilation in our toy model is shown in
Fig.~\ref{fig:spectra} for the three different final state fermion flavours;
for definiteness we assume $m_\chi=200\GeV$ and a relatively small
mass-splitting of $\mu=1.1$. The spectra of secondary photons that stem from
the subsequent decay or fragmentation of the produced fermions are derived
using \textsc{Pythia 6.4.19}~\cite{Sjostrand:2006za}. Note that in case of
bottom-quark final states we also take into account the production of VIB
gluons following Refs.~\cite{Asano:2011ik, Garny:2011ii}.\footnote{We use
throughout the values $\alpha_\text{s}=0.118$ and $\alpha_\text{em}=1/128$ as
evaluated at the mass of the $Z$ boson. For DM masses $m_\chi=40$ to $300\GeV$
this approximation affects the VIB photon cross-section at the few percent
level, and the gluon VIB cross-section by $\lesssim 20\%$.} For two-body
annihilation, we cross-checked our results with the analytical fits from
Ref.~\cite{Fornengo:2004kj, Cembranos:2010dm} and find very good agreement.
From Fig.~\ref{fig:spectra} it is clear that for small enough mass-splittings
the gamma-ray spectrum at high energies is completely dominated by VIB
photons, which show up as a pronounced peak at energies close to the dark
matter mass.  Secondary photons and FSR only become relevant at lower
energies, or for larger values of $\mu$.  In our spectral analysis of galactic
center fluxes presented in Section \ref{sec:method}, we will entirely
concentrate on the spectral VIB feature and neglect the featureless secondary
photons. We will consider the range $1<\mu\lesssim2$, because the VIB feature
is most important in the nearly degenerate case. In this range, the
\emph{shape} of the VIB spectrum is almost independent of $\mu$ (it becomes
slightly wider for larger $\mu$), but its normalization can vary rather
strongly: for $\mu=1.1$ ($\mu=2.0$), the rate is already suppressed by a
factor of 0.55 (0.05) with respect to the exactly degenerate $\mu=1$ case; for
large $\mu$, the rate scales as $\propto\mu^{-4}$ (whereas the two-body
annihilation rate scales like $\propto\mu^{-2}$). For comparison with our main
results, we will also derive limits from dwarf galaxy observations (see
Section~\ref{sec:dwarfs}); in this case we will take into account both VIB and
secondary photons.

\subsection{Connection to the MSSM} 
\label{sec:MSSM}
Before continuing, let us briefly mention the connection between our toy model
and the much more often studied case of supersymmetry.  The minimal
supersymmetric extension to the standard model (MSSM) is extremely well
motivated from a particle physics point of view---leading, in particular, to a
unification of gauge couplings and strongly mitigating fine-tuning issues in
the Higgs sector---and the stability of the lightest supersymmetric particle
(LSP) is guaranteed by the conservation of $R$-parity; if it is neutral and
weakly interacting, the LSP thus makes for an ideal DM candidate (for a
comprehensive and pedagogical primer to supersymmetry and the MSSM see
e.g.~Ref.~\cite{Martin:1997ns}).

In most cases, the lightest neutralino is the LSP, and thus a prime candidate
for WIMP DM \cite{Jungman:1995df}. It is  a linear combination of the
superpartners of the neutral components of the $U(1)\times SU(2)$ gauge as
well as  Higgs fields,
\begin{equation}
  \chi\equiv\widetilde\chi^0_1= N_{11}\widetilde B+N_{12}\widetilde W^3
  +N_{13}\widetilde H_1^0+N_{14}\widetilde H_2^0\,,
\end{equation}
and thus a Majorana fermion just like the DM particle in our toy model. As
pointed out above, the annihilation into fermion-antifermion pairs $\bar ff$
is therefore helicity suppressed in the limit of small velocities; this
helicity suppression can be lifted if an additional photon is present in the
final state \emph{and} annihilation happens via the $t$-channel exchange of a
charged particle. In the case of supersymmetry, this can only be achieved
through the corresponding left- and right-handed sfermions $\tilde f_L$ and
$\tilde f_R$ which, in the limit of vanishing $m_f$, couple to the neutralino
and fermions  as 
\begin{equation}
  {\cal L}^{\chi \tilde f f}_{\rm int}= y_L \bar \chi f_L \tilde f_L+y_R \bar
  \chi f_R \tilde f_R+{\rm h.c.}\,,
\end{equation}
where as usual $f_{R/L}\equiv\frac12(1\pm\gamma_5)f$. Compared to
Eq.~\ref{eq:singlet-eR}, the sfermions thus play exactly the same role as
$\eta$ and the main difference to our toy model is that i) there are
\emph{two} relevant scalars for \emph{each} fermion final state and that ii)
the interaction strength $y_{(R,L)}$ is no longer a free parameter but
uniquely defined by gauge symmetry,  and of course the composition of the
neutralino (see e.g.~Ref.~\cite{Haber:1984rc}): 
\begin{eqnarray}
  y_L &=& - \frac{2Q_f \mp1}{\sqrt{2}}g \tan \theta_{\rm W}  N_{11}
  \mp\frac{g}{\sqrt{2}}N_{12}\,,  \label{yl}\\
  y_R &=& \sqrt{2}Q_f g \tan\theta_{\rm W} N_{11}\,, 
\end{eqnarray}
where $g$ is the usual $SU(2)$ coupling and $\theta_{\rm W}$ the Weinberg
angle.  In Eq.~(\ref{yl}), the upper (lower) signs apply if the third
component of the weak isospin for $f$ is given by $T_3=+1/2$ ($T_3=-1/2$).

While the couplings are fixed, the relative importance of the $\bar f f\gamma$
final state to the total annihilation rate in the $v\rightarrow0$ limit
strongly depends, as we have seen, on the mass difference $\mu_{\tilde
f}\equiv(m_{\tilde f}/m_\chi)^2$ between neutralino and sfermion. In a
phenomenological description of the MSSM, i.e.~without referring to a specific
mechanism for supersymmetry breaking, all $\mu_{\tilde f}$ can be treated as
essentially free parameters; choosing one sfermion to be much more degenerate
in mass with the neutralino than all the others thus effectively, for the
purpose of our discussion, reduces the general MSSM case to our toy
model.\footnote{ Note, however, that $y$ still cannot be treated as a free
parameter. Let us also stress that even if choosing one $\mu_{\tilde f}$ so
small that the annihilation $\chi\chi\rightarrow\bar f f\gamma$ dominates the
gamma-ray spectrum at high energies, this does \emph{not} imply that this
process (or $\chi\chi\rightarrow\bar f f$) also dominates the gamma-ray
spectrum at $E_\gamma\ll m_\chi$ or is most important in setting the relic
density. The effective equivalence between our toy model and the MSSM in the
limit $\mu\rightarrow1$ thus really only refers to the form of the gamma-ray
spectrum at energies at and slightly below $m_\chi$.}

In fact, such a situation is phenomenologically very relevant already in the
most minimal supersymmetric setup, the constrained MSSM (CMSSM), and appears
both in the $\tilde\tau$- and (for neutralino masses larger than what we are
interested in here) in the $\tilde t$-coannihilation region---which apart from
the focus point, funnel and bulk region are the only regions in the CMSSM
parameter space where the neutralino acquires the correct thermal relic
density (see e.g. Ref.~\cite{Ellis:2003cw} for a discussion; note, however,
that the bulk region now is already excluded by LHC data). Correspondingly, it
was found in Ref.~\cite{Bringmann:2007nk} that VIB can indeed greatly dominate
the total photon spectrum at high energies in the MSSM in general, but in
particular in the coannihilation regions of the CMSSM.  If only the stau is
degenerate in mass with the neutralino (as is the case for example in
benchmark model BM2 of that reference), the shape of the resulting gamma-ray
spectrum at high energies is almost indistinguishable from the
$\tau^+\tau^-\gamma$ case displayed in Fig.~\ref{fig:spectra}; the same holds
for the more common situation that \emph{all} leptons are rather degenerate
with the neutralino (as in benchmark model BM3; small shape differences in the
high energy spectra can be attributed to slightly different mass splittings).
BM3 lies with a neutralino mass of $m_\chi=233.3\GeV$ in the reach of Fermi
LAT, and we will comment on this particular benchmark point in light of our
results below.

\section{Fermi LAT search for VIB signatures at the galactic center}
\label{sec:method}
The VIB gamma-ray signal produced in our  toy model is sharply peaked at
energies close to the DM mass, very much like gamma-ray lines that are
produced in the two-body annihilation into $\gamma\gamma$. For this reason,
our analysis will closely follow the spirit of traditional gamma-ray line
searches~\cite{Pullen:2006sy, Abdo:2010nc, Vertongen:2011mu}, as some of us
recently proposed in Ref.~\cite{Bringmann:2011ye}. The practical advantage of
DM signals that are very concentrated in the energy spectrum is that searches
for these signals can be restricted to relatively small energy windows, being
only a factor of a few larger than the energy resolution of the instrument.
For such small energy ranges, it is reasonable (and in fact {\it a posteriori} justified 
by our results) to approximate astrophyscial
background fluxes by a simple power-law; its normalization and spectral index
are obtained directly from a fit to the data. Since a detailed understanding
of the background sources is not necessary in this case, it is possible to
choose even complex (but very promising) regions like the Galactic center as 
target  in the search for a DM signal, as we will do in our present
analysis.

We will here concentrate on DM masses in the range $40\GeV< m_\chi < 300\GeV$.
The lower end of the mass range is motivated by the LEP constraint on the mass
of charged exotic particles, which reads $m_\eta>40\GeV$ (see
Section~\ref{sec:limits}); since we are mainly interested in the degenerate
scenario where $m_\chi\approx m_\eta$, this already excludes much lighter dark
matter particles. For DM masses above $300\GeV$, on the other hand, the
spectral feature would be outside of the nominal energy range of the Fermi
LAT. As discussed above, we concentrate here on mass splittings in the range
$1<\mu\lesssim 2$, for which the spectral shape of the VIB signal is
practically independent of $\mu$.

\subsection{Dark matter signal from the galactic halo}
The gamma-ray flux from DM annihilation in the galactic DM halo is given by a
line-of-sight integral over the DM density squared,
\begin{equation}
  \frac{dJ_\gamma}{dEd\Omega}(\xi) = \frac{\langle \sigma
  v\rangle}{8\pi \,m_\chi^2}
  \,\frac{dN}{dE} \int_\text{l.o.s.} ds
  \,\rho_\chi^2(r)\;.
  \label{eqn:fluxADM}
\end{equation}
Here, $m_\chi$ is the DM mass, $\langle \sigma v\rangle$ the total DM
annihilation cross-section averaged along the line of sight, $dN/dE$ the
energy spectrum of produced gamma rays, and $\xi$ denotes the angle to the
Galactic center.  The coordinate $s\geq0$ runs along the line of sight, and
the distance to the Galactic center $r$ is given by $r(s,\xi) =
\sqrt{(r_0-s\cos\xi)^2 + (s\sin\xi)^2}$, where $r_0=8.5\kpc$ denotes the
distance between Sun and the Galactic center.

We will consider the following generalized Navarro-Frenk-White
(NFW~\cite{Navarro:1996gj, Abdo:2010nc}) profile
\begin{align}
  \rho_\chi(r) \propto \frac{1}{(r/r_s)^\alpha
  \left(1+r/r_s\right)^{3-\alpha}}\;,
  \label{}
\end{align}
normalized to the fiducial value $\rho_\chi=0.4\GeV\cm^{-3}$ at Sun's
position~\cite{Catena:2009mf, Salucci:2010qr} and with a scaling radius of
$r_s=20\kpc$. In case of an inner slope of $\alpha=1$ this reproduces the
standard NFW profile. The possible impact of adiabatic
contraction~\cite{Blumenthal:1985qy, Gnedin:2003rj, Gustafsson:2006gr,
Gnedin:2011uj} can be studied in an effective way by allowing for larger inner
slopes of the profile. We will concentrate on the range $1<\alpha\lesssim1.4$,
which is still compatible with microlensing and dynamical
observations~\cite{Iocco:2011jz} (traditional adiabatic contraction following
Ref.~\cite{Blumenthal:1985qy} would give rise to even larger values of
$\alpha$, see \fex Ref.~\cite{Iocco:2011jz}). The modified isothermal and
Einasto profiles~\cite{Navarro:2003ew, Springel:2008cc, Pieri:2009je} are
expected to give comparable results to the standard NFW profile in searches
for line-like features~\cite{Vertongen:2011mu}.

\subsection{Event and target region selection}
\begin{figure}[t]
  \begin{center}
    \includegraphics[width=0.45\linewidth]{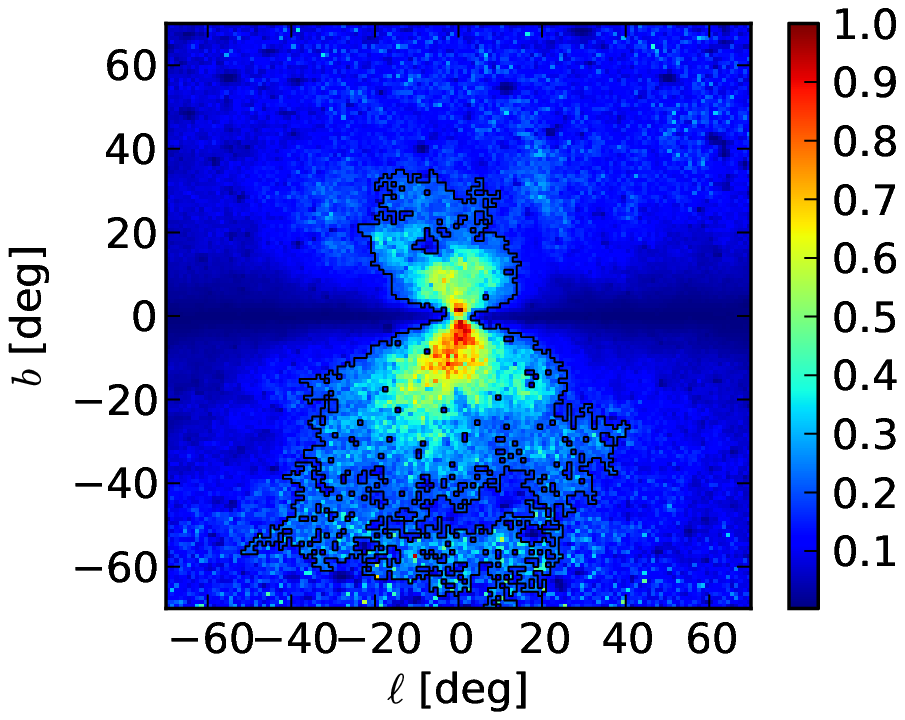}
    \includegraphics[width=0.45\linewidth]{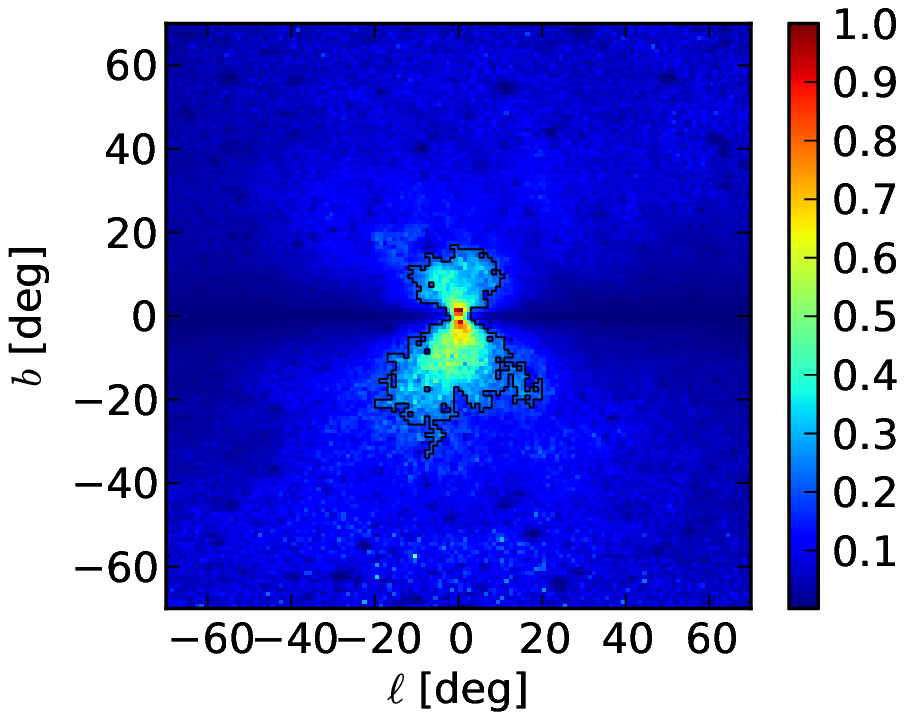}
    \includegraphics[width=0.45\linewidth]{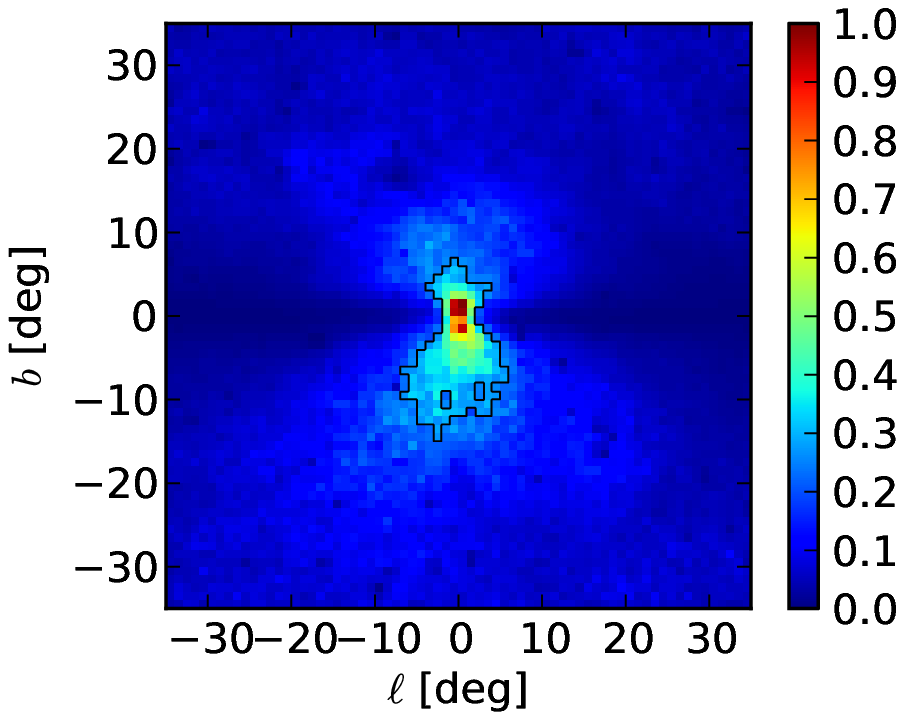}
    \includegraphics[width=0.45\linewidth]{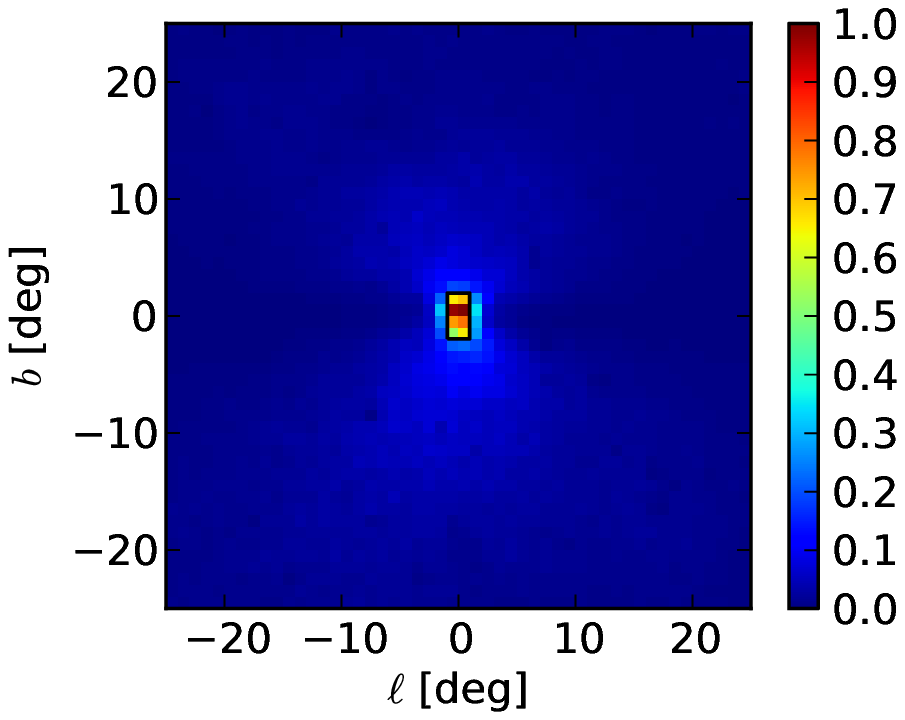}
    \vspace{-0.5cm}
  \end{center}
  \caption{Target regions used in our spectral analysis (solid black lines). From top left to
  bottom right, Reg1, Reg2, Reg3 and Reg4 are respectively optimized for DM
  profiles with inner slopes of $\alpha=(1.0,1.1,1.2,1.4)$ as described in the
  text and in Appendix \ref{apx:targetRegions}. The optimization maximizes the
  signal-to-\emph{noise} ratio. For comparison, the colors show the expected
  signal-to-\emph{background} ratio, normalized in each case to 1 for the central pixel.}
  \label{fig:targetRegions}
\end{figure}
The gamma-ray data measured by the Fermi LAT is publicly
available~\cite{Fermi:Website}. The events that enter our main analysis are
taken from the \textsf{P7CLEAN\_V6} event class.  We consider 43 months of
data (from 4 Aug 2008 to 6 Feb 2012), and select front- and back-converted
events with energies in the range 1--300 GeV. We apply a zenith angle cut of
$\theta<100^\circ$ in order to avoid contamination with photons from the earth
albedo, as well as the quality cut filter \textsf{DATA\_QUAL==1} (all event
selection is done using the 06/10/2011 version of \textsf{ScienceTools
v9r23p1}.

The  \emph{target region} for our spectral analysis is chosen by
using a data-driven adaptive procedure with the aim of maximizing the expected
\emph{signal-to-noise} ratio. We stress that such an approach is extremely
important when looking for spectral features at the statistical limit of the
detector, and that  an inefficiently chosen target region can easily wash
out or hide a potential signal.  Our choice of the optimal target region
depends on the adopted DM profile; in order to select it, we estimate the
expected spatial distribution of background noise in our search for spectral
features above $40\GeV$ by considering the actually measured events
\emph{below} $40\GeV$.  The spatial distribution of signal photons, on the
other hand, just follows from Eq.~\eqref{eqn:fluxADM}. All details of the
method are given in Appendix \ref{apx:targetRegions}.

We adopt four reference values for the inner slope of the DM profile,
$\alpha=1.0$, $1.1$, $1.2$ and $1.4$, for which we obtain the target regions
that are shown in Fig.~\ref{fig:targetRegions} as solid black lines. In this
plot, the colors encode the expected signal-to-background ratio in different
regions of the sky, normalized to one for the pixel where this ratio is
maximal (note that the actual value of this quantity is a factor of 1.9 (3.9,
31) larger for Reg2 (Reg3, Reg4) than for Reg1).  In case of a standard NFW
profile with $\alpha=1.0$, the target region includes besides the galactic
center also regions at higher and lower latitudes up to $|b|\lesssim
70^\circ$; for steeper profiles the optimal target regions shrink drastically
to regions closer to the galactic center. The galactic disc is strongly
disfavoured in all cases. Southern regions are somewhat preferred, since the
diffuse gamma-ray emission from our galaxy is not perfectly north/south
symmetric. From these four regions we extract the measured spatially
integrated gamma-ray energy spectrum for our subsequent analysis.

\subsection{Spectral analysis}
\label{sec:specan}
The search for VIB signatures is done by using the \emph{sliding energy
window} technique discussed  \fex in Refs.~\cite{Pullen:2006sy, Abdo:2010nc,
Vertongen:2011mu, Bringmann:2011ye}: we consider for each DM mass $m_\chi$
 in the range $40\GeV < m_\chi < 300 \GeV$ a small energy window that is
approximately centered on $m_\chi$, and hence on the position of the expected
VIB feature. More precisely, we use the energy range
$E=m_\chi\epsilon^{-0.7}$\dots$\min[m_\chi\epsilon^{0.3},300\,{\rm GeV}]$, where
the size of the window $\epsilon$ varies between $\epsilon \simeq
1.8$ for $m_\chi=40\GeV$ and $\epsilon\simeq7$ for $m_\chi=300\GeV$. The
size of the window is identical to the values used in
Ref.~\cite{Vertongen:2011mu}, where it was found to lead to reasonable
background fits in context of gamma-ray line searches. The position of the
window is not exactly symmetric around $m_\chi$, but slightly shifted towards
lower energies as it was suggested for VIB features in
Ref.~\cite{Bringmann:2011ye} in order to increase the sensitivity. We
emphasize again that secondary photons, as they come from the decay or
fragmentation of the fermions, are neglected in our spectral analysis, because
they become relevant only \emph{outside} of the energy window that we consider 
here (at least for our toy model).

For each given mass $m_\chi$, and within the adopted small energy window, we
now fit the gamma-ray spectra measured in the different target regions of
Fig.~\ref{fig:targetRegions} with a simple three-parameter model: The
astrophysical background fluxes are approximated by a power law with a free
spectral index (1) and normalization (2); the DM VIB signal has only a free
normalization (3), whereas its mass and the mass-splitting (which we set to
$\mu=1.1$ in most of the analysis) remains fixed during the fit. For physical
reasons we require the normalization of the VIB signal to be positive.

Technically, we perform a binned analysis of the gamma-ray spectrum measured
in the different target regions. To this end, we distribute the corresponding
measured gamma-ray events in a very large number of energy bins (200 per
decade). Since the size of the adopted energy bins is much smaller than the
energy resolution of the Fermi LAT, our analysis is essentially identical to
an unbinned analysis of the energy spectrum. Each energy bin $j$ then contains
a number $c_j$ of events. The number of expected events $\mu_j$ are obtained
by convolving our above three-parameter model with the energy dispersion and
the exposure of the LAT; the resulting $\mu_j$ can then be fitted to the
observed counts $c_j$ by maximizing the likelihood function $\mathcal{L} =
\Pi_j P(c_j|\mu_j)$ with respect to the three model parameters.  Here,
$P(c|\mu)$ is the Poisson probability to observe $c$ events when $\mu$ are
expected.  Note that the functional form of the energy dispersion is directly
inferred from the IRF of the \textsf{P7CLEAN\_V6} event class and correctly
averaged over impact angles and front- and back-converted events, using our
own software. Exposure maps are derived using the \textsf{ScienceTools
v9r23p1}. 

Limits on or the significance of a DM VIB contribution can then be derived by
using the profile likelihood method~\cite{Rolke:2004mj}. A one-sided $95\%$
C.L.~upper limit on the annihilation cross-section is obtained by increasing
the DM signal normalization from its best-fit value until $-2\ln\mathcal{L}$
increased by a value of 2.71 (while refitting the background parameters). The
significance of a signal, on the other hand, is derived from the test
statistics (TS)
\begin{align}
  TS = -2\ln\frac{\mathcal{L}_\text{null}}{\mathcal{L}_\text{best-fit}}\;,
  \label{eqn:TS}
\end{align}
where $\mathcal{L}_\text{best-fit}$ is the likelihood of the best-fit model,
and $\mathcal{L}_\text{null}$ the likelihood of the null hypothesis (the
absence of a DM signal; the null model has only two free parameters).  In
absence of a signal, one expects that the TS follows some $\chi^2$
distribution.  More precisely, since the normalization of the DM signal is
bounded to be positive, the TS should follow a
$0.5\chi^2_{k=0}+0.5\chi^2_{k=1}$ distribution (see \fex
Ref.~\cite{Vertongen:2011mu}), where $\chi_{k=0}^2$ and $\chi^2_{k=1}$ have
zero and one degree of freedom, respectively.\footnote{The probability
distribution function of $\chi^2_{k=0}$ is just $\delta(TS)$. For discussions
about the coverage of confidence intervals on bounded parameters see
Ref.~\cite{Rolke:2004mj}.} This theoretical expectation is indeed very well
confirmed by a subsampling analysis of the data as we discuss in
Appendix \ref{apx:details}. Hence, if the test statistics is measured to be
$TS$ for a certain DM mass in a single trial, this would correspond to
slightly more than $\sqrt{TS}\sigma$ significance. 

However, since in the present analysis we effectively perform many
statistically independent trials when scanning through $m_\chi$ and analyzing
different target regions, the probability to find just by chance a statistical
fluctuation that mimics a signal is increased; this is known as the
look-elsewhere effect (LEE). In our case, we approximate the distribution of
maximal TS values from which the significance of our signature is calculated
by $4\times4=16$ trials over a $\chi_{k=2}^2$ distribution. Four trials over a
$\chi_{k=2}^2$ distribution come from the scan over $m_\chi$ (see Appendix
\ref{apx:details} or, for a
general discussion,   Ref.~\cite{Gross:2010qma}); the remaining four trials
are associated with the four target regions. In practice, the significance of
the observed signature is then found by solving
$P(\chi^2_{k=2}<TS)^\text{\#trials}=P(\chi^2_{k=1}<\sigma^2)$ for $\sigma$,
where $P(\chi_{k}^2 < x)$ denotes the probability to observe a value smaller
than $x$ when drawing from a $\chi^2_k$ distribution.

\subsection{Results}
\begin{figure}[t]
  \begin{center}
    \includegraphics[width=0.7\linewidth]{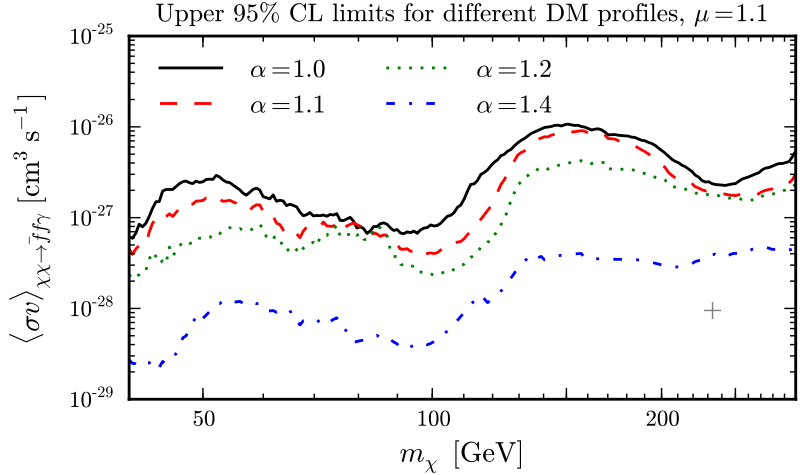}
    \vspace{-0.7cm}
  \end{center}
  \caption{Our results for the 95\% CL limits on the \emph{three-body} (VIB)
  annihilation cross-section of $\chi\chi\to \bar f f\gamma$, for different
  values of the DM profile inner slope $\alpha$. The limits are obtained by a
  spectral analysis of the gamma-ray flux in the corresponding target regions
  shown in Fig.~\ref{fig:targetRegions}. We assume a mass splitting of
  $\mu=1.1$. Note that the limits do not directly depend on the nature of the
  final state fermion $f$, as they are derived from the VIB feature
  \emph{only}. The gray cross shows the CMSSM benchmark point BM3 from
  Ref.~\cite{Bringmann:2007nk}.}
  \label{fig:vibBoosts}
\end{figure}

In our analysis, no VIB signal with a significance of at least $5\sigma$ were found.
Instead, we show in Fig.~\ref{fig:vibBoosts} the \emph{upper limits} at 95\% CL on 
the VIB three-body cross-section, for the different values
of the inner slope $\alpha$ that correspond to the target regions in
Fig.~\ref{fig:targetRegions}. We assumed a mass-splitting of $\mu=1.1$ for
definiteness; limits for other values of $\mu$ are very similar and will be
discussed below (see \fex Fig.~\ref{fig:vibMUMU}). As can be seen from the
plot, our limits are always stronger than the 'thermal cross-section' of
$3\times10^{-26}\cm^3\s^{-1}$ that is often quoted for comparison; in the case
of contracted profiles with $\alpha=1.4$ they can even reach down to values of
$10^{-28}\cm^3\s^{-1}$ for DM masses $m_\chi\lesssim100\GeV$. As we will
discuss below in Section~\ref{sec:dwarfs}, our limits are much stronger than
what can be obtained from \fex dwarf galaxy observations. For comparison, the
gray cross in Fig.~\ref{fig:vibBoosts} shows the CMSSM benchmark point
BM3~\cite{Bringmann:2007nk}, which lies in the coannihilation region and was
already discussed above. This benchmark point still remains unconstrained by
our limits; its rather small cross-section is closely related to the
requirement that the neutralino is a thermal relic, as we will discuss in
Section \ref{sec:thermal} below.\medskip

\begin{figure}[t]
  \begin{center}
    \includegraphics[width=0.545\linewidth]{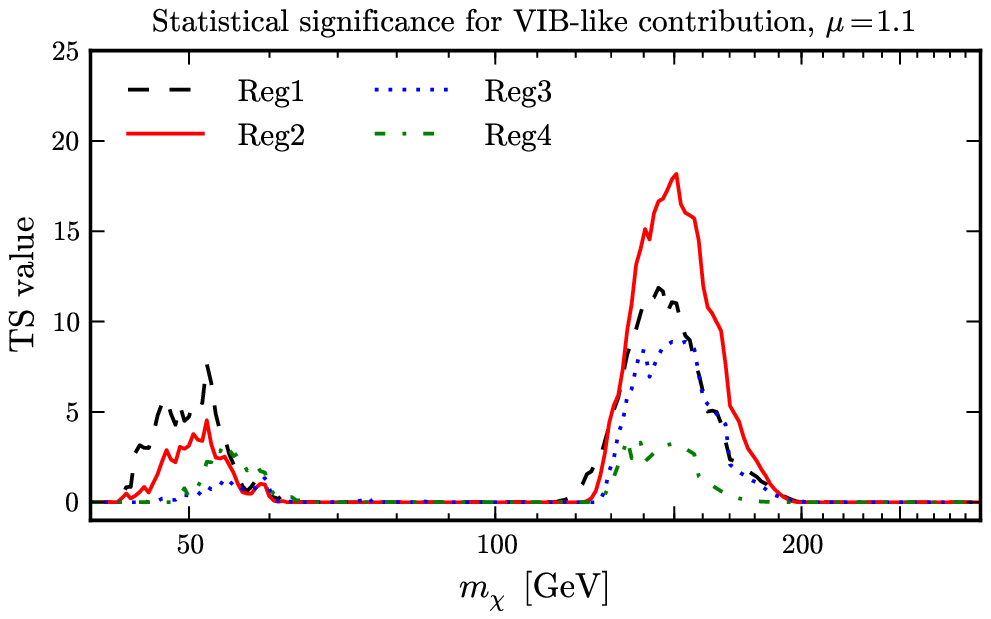}
    \includegraphics[width=0.445\linewidth]{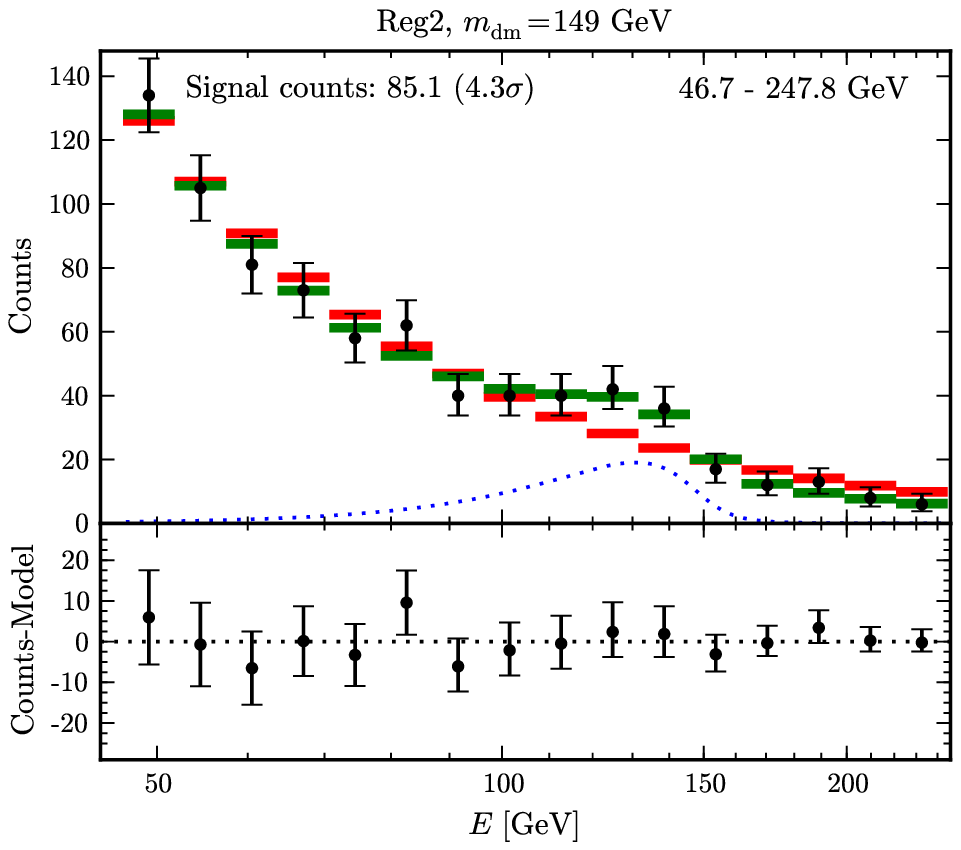}
    \vspace{-1.1cm}
  \end{center}
  \caption{Left panel: statistical significance for VIB-signal in terms of the
  TS value, as function of $m_\chi$ and for the different target regions shown
  in Fig.~\ref{fig:targetRegions}. Right panel: fits to data in Reg2 for the
  best signal candidate at $m_\chi=149\GeV$. We show the background-only fit
  without DM signal as red bars. The green bars show the background plus DM
  signal fit, the blue line the corresponding VIB signal flux. In the
  right panel, we rebinned the data into (9 times) fewer bins than actually
  used in our statistical analysis in order to improve the optical appearance
  of the figure. Note that the shown fluxes are already integrated over the
  individual energy bins and properly convolved with the LAT IRF.}
  \label{fig:fit}
\end{figure}

In the left panel of Fig.~\ref{fig:fit}, we show the \emph{significance} for a
VIB-like spectrum as function of $m_\chi$, assuming that $\mu=1.1$. The
different lines correspond to the different target regions. The significance
is shown in terms of the TS value that was discussed above. We find a possible
signal candidate at a DM mass of $m_\chi\approx 150\GeV$.  The indication for
a signal is largest for the target region Reg2, which corresponds to
$\alpha=1.1$, and has a nominal significance of $\sqrt{TS}=4.3\sigma$. Taking
into account the LEE as discussed above, the significance is $3.1\sigma$. The
corresponding fit to the data is shown in the right panel of
Fig.~\ref{fig:fit}; the spectral feature in the measured flux can be easily
recognized by eye. A similar preference for a signal, although with less
significance, appears also in the other regions Reg1, Reg3 and Reg4 (note that
the fluctuations around 50 GeV are completely within the statistical
expectations). TS values of zero indicate that for these values of $m_\chi$
the data would be best fitted with an unphysical negative signal
normalization; in this case, the likelihood of the model with DM contribution
becomes identical to that of the null model because  we enforced a non-negative signal
normalization in our fits.

We have performed several tests to exclude the tempting DM interpretation of
this signature, none of which has succeeded so far: By masking out different
halves of the signal region of Reg2, for example, we find that the signal
independently appears  in the north, south, east and west parts of Reg2
(though with a large scatter in the significances), as expected from a DM
signal. When shifting the target region away from its position by about
$10$--$20^\circ$, on the other hand, the signal disappears completely.  This
makes a purely instrumental effect, which would likely also appear in other
sky regions, less likely. To test our statistical method, we performed a
subsampling analysis of the Fermi LAT data in the galactic
anticenter region in order to obtain an empirical understanding of the
statistical behaviour of our TS (see Appendix \ref{apx:details} for details).
We find that---in absence of a signal---our TS follows very well a $\chi^2$
distribution, as theoretically expected. The signature appears also in the
\textsc{P7ULTRACLEAN\_V6} and \textsc{P7SOURCE\_V6} event classes, and when
considering back- and front-converted events separately; furthermore, as
expected for a real signal, its significance has grown with time (i.e.~we find
smaller $TS$ values when considering instead older and older data sets, though
with some scatter around the linear trend).  The uncertainty in the effective
area as relevant for searches for line-like features is about
$2\%$~\cite{Fermi:caveatsPASS7}, which is much smaller than the fractional
contribution of the observed signature to the flux.

Despite these encouraging facts, we call for caution when interpreting this
signature as due to DM annihilation. First of all, with a significance of
$3.1\sigma$ (when taking the LEE into account), it could still simply be an
upward fluctuation at the right place.  Alternatively, the observed signal
could be real but due to a yet unidentified astrophysical process, like \fex
the inverse Compton scattering of an extremely hard electron component on
stellar light (see \fex Ref.~\cite{Su:2010qj} for a related discussion of ICS
in the context of the Fermi bubbles)---though in general, as already stressed
several times, it is quite difficult to explain this kind of spectral features
with astrophysical processes. In this context, it is also worth to mention
again that our analysis crucially depends on the commonly adopted assumption
that the background {\it locally} follows a power-law, i.e.~within each energy
window that we consider. In principle, it might thus make for an interesting
follow-up study to perform a signal fit on a more complicated background which
contains, e.g., a break in the spectral index that could be confused with a
signal.  We do not expect, however, that such an analysis leads to
qualitatively different results than presented here because the data itself
tells us that the single power-law assumption works very well (see Appendix
\ref{apx:details}), and because of the sharpness of the observed signature.
Even more importantly, we always find a spectral index for the background
contribution that is consistent with $-2.6$; this value is expected for the production of
gamma rays from the collision of cosmic rays with the interstellar medium and
thus extremely well motivated from astrophysics  (see \fex
Ref.~\cite{FermiLAT:2012aa}). Lastly, our analysis relies
entirely on the publicly available data, which makes it impossible to take
into account all known instrumental effects.  However, we strongly believe
that our actual \emph{statistical} analysis is sound, and not significantly
affected by the obvious systematics of the LAT. 

Finally, we note that the best-fit three-body cross-section for a VIB signal
in Reg2 is $\langle \sigma v\rangle=(6.2\pm 1.5\
^{+0.9}_{-1.4})\times10^{-27}\cm^3\s^{-1}$ (assuming $\alpha=1.1$), the
best-fit mass is in the range $m_\chi=149\pm4\ ^{+8}_{-15}\GeV$; the errors
are respectively statistical and systematical.\footnote{Systematical errors
stem from uncertainties in the overall effective area ($10\%$) and in the
energy calibration ($^{+5}_{-10}\%$), see Ref.~\cite{Fermi:caveatsPASS7}.}
This is quite a bit larger than what is actually expected for VIB  from
\emph{thermally} produced DM (see Section~\ref{sec:thermal} for a discussion),
making a straightforward interpretation of the signal in terms of a VIB signal
somewhat less appealing.  However, let us point out that the signature can
also well be fitted by a pure gamma-ray line as produced in
$\chi\chi\to\gamma\gamma$, with a dark matter mass of around
$m_\chi\simeq130\GeV$ and an annihilation cross-section $\langle\sigma
v\rangle \sim 10^{-27}\cm^3\s^{-1}$ (see also Ref.~\cite{Weniger:2012tx} for a
dedicated analysis). Finally, we note that the more commonly adopted
annihilation spectra from e.g.~$\chi\chi\to\mu^+\mu^-$ or $\chi\chi\to\bar bb$
annihilation are much too flat to fit the data and thus cannot be used to
explain the signal.

\section{Comparison with other constraints}
\label{sec:constraints}
The limits on the cross-section $\langle\sigma v\rangle_\text{3-body}$ as
shown in Fig.~\ref{fig:vibBoosts} were derived from VIB photons \textit{only},
according to the spectrum given in Eq.~\eqref{eqn:sv3spec}. We will now
discuss these constraints in light of other complementary probes, namely
limits from gamma-ray observations of dwarf galaxies, limits that can be
derived from the thermal production of DM and collider limits; we finally
stress that small values of $\mu$, for the annihilation into $\bar b b\gamma$
final states, can also very efficiently be probed by both cosmic ray
antiprotons and direct DM detection experiments. 

\begin{figure}[t]
  \begin{center}
    \includegraphics[width=0.7\linewidth]{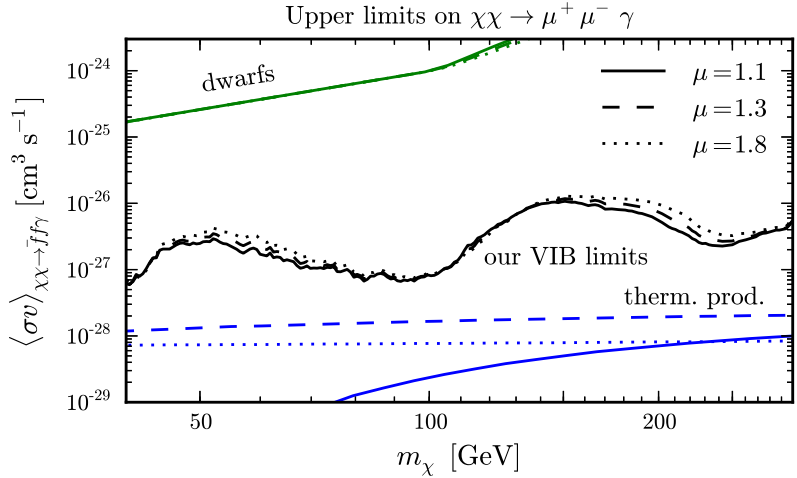}
    \vspace{-0.7cm}
  \end{center}
  \caption{Comparison of different upper limits on the three-body annihilation
  cross-section of $\chi\chi\to \mu^+\mu^-\gamma$, for three reference values
  of the mass splitting $\mu$. Black lines show 95\% CL upper limits that come
  from our spectral analysis of the Galactic center fluxes as shown in
  Fig.~\ref{fig:vibBoosts}, assuming a standard NFW profile. Green lines
  (partially overlapping) show the corresponding limits derived from dwarf
  galaxy observations, taking into account both two- and three-body
  annihilation channels. Blue lines show upper limits on the annihilation
  cross-section that are derived from requiring that the relic density
  predicted by our toy model does not \emph{undershoot} the observed value.}
  \label{fig:vibMUMU}
\end{figure}

\begin{figure}[t]
  \begin{center}
    \includegraphics[width=0.7\linewidth]{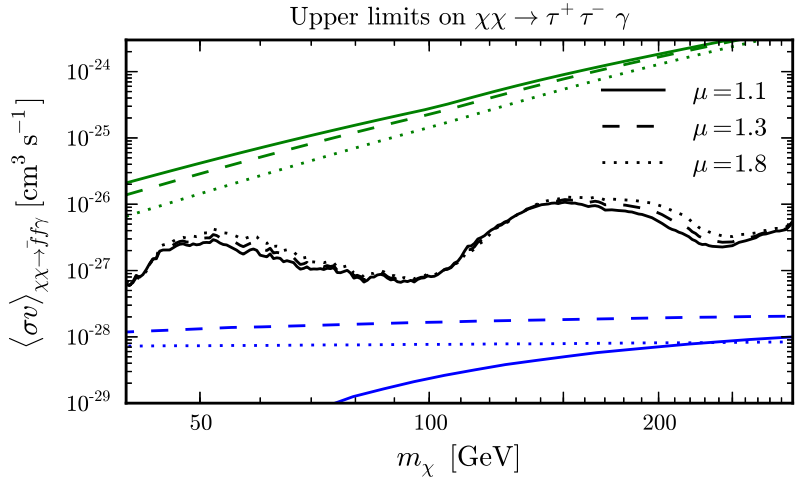}
    \vspace{-0.7cm}
  \end{center}
  \caption{Same as Fig.~\ref{fig:vibMUMU}, but for annihilation into
  $\tau^+\tau^-\gamma$.}
  \label{fig:vibTAUTAU}
\end{figure}

\begin{figure}[t]
  \begin{center}
    \includegraphics[width=0.7\linewidth]{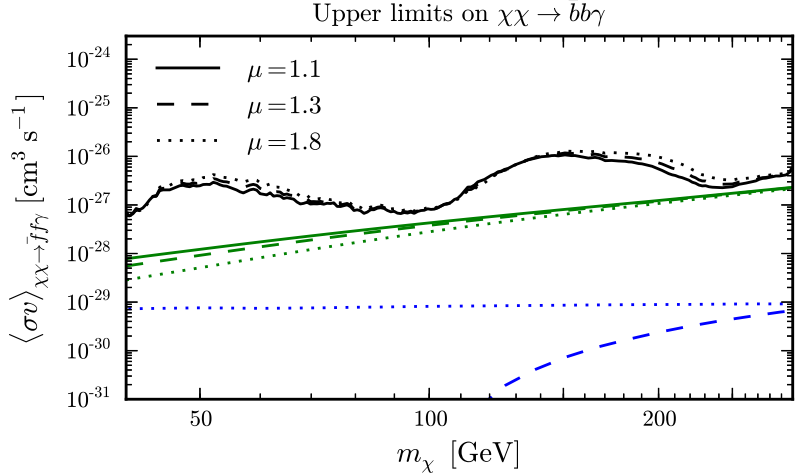}
    \vspace{-0.7cm}
  \end{center}
  \caption{Same as Fig.~\ref{fig:vibMUMU}, but for annihilation into
  $\bar{b}b\gamma$. Note that we include $\bar{b}bg$ final states when
  calculating the dwarf limits~\cite{Asano:2011ik, Garny:2011ii}.}
  \label{fig:vibBBAR}
\end{figure}

\subsection{Limits from dwarf galaxies} 
\label{sec:dwarfs}
Important limits on the DM annihilation cross-section derive from gamma-ray
observations of dwarf galaxies~\cite{Abdo:2010ex, Ackermann:2011wa,
GeringerSameth:2011iw}. To compare our above results in
Fig.~\ref{fig:vibBoosts} with the constraints that typically can be obtained
from these targets, we will make use of the flux limit presented in
Ref.~\cite{GeringerSameth:2011iw}.\footnote{ Refs.~\cite{Abdo:2010ex,
Ackermann:2011wa} present limits on different common annihilation channels
that are however difficult to translate into limits on VIB.} It reads, in
terms of the total annihilation cross-section, $\langle\sigma v\rangle<8\pi\
m_\chi^2/ N_\gamma^\text{tot}\cdot5.0\times10^{-30}\cm^3\s^{-1}\GeV^{-2}$.
Here, $N_\gamma^\text{tot}$ denotes the number of photons that are produced
per annihilation in the energy range $1$--$100\GeV$.  When calculating
$N^\text{tot}_\gamma$, we fully take into account all photons from three-body
as well as the common two-body final states according to Eqs.~\eqref{eqn:sv2s}
and \eqref{eqn:sv3};\footnote{For simplicity, we here approximated the
spectrum of secondary photons from the VIB process with the secondary photon
spectrum of the corresponding two-body process. This makes the dwarf limits
slightly too strong at high DM masses (by less than a factor of two).} the
relative importance of two- and three-body final states depends on $\mu$ and
the mass of the final state fermions.

We plot the resulting limits in Figs.~\ref{fig:vibMUMU}, \ref{fig:vibTAUTAU}
and \ref{fig:vibBBAR} as green lines for different values of the mass
splitting $\mu$ and for the three different final state fermion flavours; in
order to allow a simple comparison with our result in
Fig.~\ref{fig:vibBoosts}, we choose to present them in terms of the
\emph{three-body} annihilation cross-section (by rescaling them by a factor of
$\langle \sigma v\rangle_\text{3-body}/[\langle \sigma v\rangle_\text{2-body}
+\langle \sigma v\rangle_\text{3-body}] $).  In most cases the limits depend
relatively strongly on the mass splitting $\mu$ (Figs.~\ref{fig:vibTAUTAU} and
\ref{fig:vibBBAR}); only for annihilation into $\mu^+\mu^-$ the limits remain
practically independent of $\mu$, since the fluxes are always dominated by VIB
photons (\cf Figs.~\ref{fig:spectra} and~\ref{fig:vibMUMU}). In contrast to
that, our limits from the spectral search for VIB features in the galactic
center, as shown by the black lines, exhibit only a very weak dependence on
the mass splitting parameter $\mu$ for values of $\mu\approx1.1$--$2$, since
the spectral shape of the VIB radiation changes only mildly in this range.

As long as the mass splitting $\mu$ remains small, and for the leptonic final
states, our spectral search in the galactic center fluxes for VIB features
leads to much \emph{stronger} constraints on our toy model than dwarf galaxy
observations, see Figs.~\ref{fig:vibMUMU} and~\ref{fig:vibTAUTAU}. For a large
enough mass splitting, or in case of colored final state fermions like
bottom-quarks, however, the 3-body annihilation into VIB photons becomes less
relevant and the dwarf limits start to overtake our VIB limits (\cf
Fig.~\ref{fig:vibBBAR}). Note that for colored final states the VIB of gluons
becomes important~\cite{Garny:2011ii, Asano:2011ik} and actually dominates the
dwarf limits (see also Section~\ref{sec:limits2}).

\subsection{Thermal production}
\label{sec:thermal}
In our toy model, self-annihilation in the early Universe is usually dominated
by the $p$-wave process $\chi\chi\to \bar ff$, since the averaged velocity is
$v\sim\mathcal{O}(1)$, see Eq.~\eqref{eqn:sv2p}. Any embedding of this model
in a full theory (like the MSSM, see Section~\ref{sec:MSSM}) would likely
increase the total annihilation rate. A loose but reasonable \emph{upper
limit} on the coupling constant $y$ can hence be derived by requiring that the
relic density which follows from the $t$-channel process $\chi\chi\to \bar ff$
alone does not already undershoot the measured DM density. From this, we
could estimate an upper limit on the partial annihilation cross-section for
$\chi\chi\to\bar{f}f\gamma$ as follows: At the freeze-out temperature 
$T_f\sim m_\chi/20$, the $p$-wave annihilation channel $\chi\chi\to \bar ff$ 
is mildly suppressed by a velocity factor of $\langle v^2\rangle= 6T_f/m_\chi\sim 0.3$; 
the corresponding annihilation cross-section should be
equal to or smaller than the thermal cross-section $\langle\sigma
v\rangle_\text{th}\simeq3\times 10^{-26}\cm^3\s^{-1}$, in order to not
undershoot the observed DM relic density. However, the $s$-wave VIB process
$\chi\chi\to \bar ff\gamma$ is suppressed by a factor
$\sim\alpha_\text{em}/\pi$ (for final state fermions with charge one), 
independently of the temperature $T$ or the velocity $v$.  Taking these pieces 
of information together, one finds an upper limit of roughly 
$\frac{\alpha_\text{em}}{0.3\pi}\langle\sigma v\rangle_\text{th}\sim 
2\times10^{-28}\cm^3\s^{-1}$ for the three-body annihilation cross-section 
$\langle \sigma v\rangle_\text{3-body}$. Hence, if we want our toy model to be 
compatible with the observed relic density (assuming thermal DM production 
and a standard thermal history of the universe), this suggests an
annihilation cross-section $\langle\sigma v\rangle_\text{3-body}$ that is
about two orders of magnitude smaller than the common thermal cross-section.
Note, however, that \fex destructive interference with $s$-channel diagrams,
which do not appear in our toy model, could spoil this argument and thus in
principle allow for larger rates.

Neglecting the
coannihilation of $\chi$ and $\eta$,  the relic density due to two-body
annihilation $\chi\chi\to \bar ff$ would be approximately given by (see \fex
Ref.~\cite{Garny:2011cj})
\begin{align}
  \Omega_\chi h^2
  \simeq 0.11 \frac{1}{N_c}\left( \frac{0.35}{y} \right)^4
  \left( \frac{m_\chi}{100\GeV} \right)^2
  \frac{(1+\mu)^4}{1+\mu^2}\;.
  \label{eqn:RelicDensity}
\end{align}
For comparison, we performed a full calculation with
\textsc{MicrOMEGAs}~\cite{Belanger:2010gh}.
We find that our order-of-magnitude estimate is only approximately valid if
$\mu\gtrsim 1.3$ in case of leptonic final states, and if $\mu\gtrsim 1.8$ in
case of bottom quark final states (see also Ref.~\cite{Griest:1990kh}).  For
smaller mass splittings, coannihilations play a significant role.  Note
furthermore that even a vanishingly small coupling $y$ may in fact not 
sufficiently suppress the {\it total} effective annihilation rate, for low dark matter 
masses $m_\chi$, to reproduce the observed relic density simply because the 
gauge-boson mediated annihilation of $\eta\eta$ pairs will at some point start to 
dominate. This effect is most pronounced for small mass differences $\mu$ and 
can clearly be seen in Figs.~\ref{fig:vibMUMU}-\ref{fig:vibBBAR} (see also 
Ref.~\cite{Garny:2013ama}).
Furthermore, the coupling to the Higgs can become relevant at masses around 50
GeV.

Using our full numerical results, and requiring now that $\Omega_\chi h^2\gtrsim 0.1$, 
we obtain upper limits on the
cross-section as shown in Figs.~\ref{fig:vibMUMU}-\ref{fig:vibBBAR} by the blue lines for
different values of $\mu$. These limits largely agree with the rough estimates
given in the preceding paragraph
as long as the mass-splitting is $\mu\sim 1.5$. For larger mass splittings,
or for smaller splittings where coannihilation would further
reduce the relic density, the limits become stronger. 

From Figs.~\ref{fig:vibMUMU}--\ref{fig:vibBBAR} it is apparent that in case of
a regular NFW profile with ($\alpha=1$) our above galactic center limits are
still around one (two) orders of magnitude away from values of the three-body
cross-section that are consistent with a thermal relic in case of lepton
(bottom quark) final states. However, in case of
compressed profiles with $\alpha>1$, our limits become stronger, as shown in
Fig.~\ref{fig:vibBoosts}, and for inner slopes of the DM halo
$\alpha\approx1.2$--$1.4$, they can become sensitive enough to probe the
required cross-sections of $\sim 10^{-28}\cm^3\s^{-1}$ and below.

Further enhancements of the signal, without violating our relic density
constraints, can appear due to the gravitational clustering of DM as predicted 
by $N$-body simulations~\cite{Bergstrom:1998jj, Pieri:2009je}. Moreover, 
the effect of coannihilation could actually be inversed and \emph{increase} the
relic density in some cases~\cite{Profumo:2006bx}, which again would allow
larger annihilation rates while still being compatible with the relic density.
Finally, we note that the non-thermal production of dark matter or a
non-standard freeze-out history would invalidate our relic density limits and
admit larger annihilation cross-sections today (see \fex
Refs.~\cite{Moroi:1999zb, Gelmini:2006pw, Fairbairn:2008fb}).

\subsection{Limits on light charged or colored scalar particles}
\label{sec:limits}

Scenarios with light scalars that couple to the Standard Model quarks and
leptons are constrained by direct searches at colliders as well as by indirect
searches through their effect on the quantum corrections to the electroweak
precision measurements. 
\medskip

\emph{Collider searches.} The most severe constraint on scenarios with light
charged particles follows from the non-observation of an excess in the
invisible decay width of the $Z$ boson at LEP,  $\Delta \Gamma_{\rm inv}<2.0$
MeV~\cite{:2005ema}, which excludes the existence of exotic charged scalar
particles with mass below 40 GeV~\cite{Nakamura:2010zzi}. Since we are mostly
interested in the degenerate case where $m_\chi\approx m_\eta$, we only
considered DM masses with $m_\chi>40\GeV$ in this work.

Additional constraints on the mass of \emph{charged scalar particles} can be
derived from the non-observation at LEP of an excess over the Standard Model
expectations of dilepton events with missing energy, generated by the
production of charged scalar particles that decay into a lepton and an
invisible particle (in our framework, the DM particle).  The search strategy
relies on the identification of at least one lepton event as well as on
satisfying the requirements on isolation and transverse momentum. Hence, the
efficiency of the search decreases drastically when the DM particle and the
intermediate scalar are more and more degenerate in mass, which is exactly the
region we are interested in. The best constraints on such a scenario were
published by the OPAL collaboration using 680 pb$^{-1}$ of $e^+e^-$ collisions
at center-of-mass energies between 192 GeV and 209 GeV; they
read~\cite{Abbiendi:2003ji}
\begin{itemize}
  \item $m_{\eta}\geq 94.0$ GeV at the 95\% CL for couplings to muons, assuming 
    $m_{\eta}-m_\chi>4$ GeV  and 100\% branching ratio for 
    $\eta\rightarrow \mu \chi$
  \item $m_{\eta}\geq 89.8$ GeV at the 95\% CL for coupling to taus, assuming 
    $m_{\eta}-m_\chi>8$ GeV and 100\% branching ratio for 
    $\eta\rightarrow \tau \chi$.
\end{itemize}
For $\mu\approx1.1$, these limits are always satisfied by our toy model.
However, much larger mass splittings are already partially ruled if the dark
matter mass is below $\sim 90\GeV$.

Searches for light \emph{colored scalar particles} have been undertaken both
at LEP, Tevatron and at the LHC. The non-observation of an excess over the
Standard Model background of dijet events with missing energy translates into
the following 95\% CL limits on the mass of colored scalars, assuming 100\%
branching ratio for $\eta\rightarrow b \chi$ (see also Fig.~4 in 
Ref.~\cite{Asano:2011ik} for an updated summary):
\begin{itemize}
  \item  $m_{\eta}\geq 875~{\rm GeV}$, assuming $m_{\eta}-m_\chi>130~{\rm
    GeV}$ (ATLAS~\cite{Aad:2011ib})
  \item  $m_{\eta}\geq 76~{\rm GeV}$, assuming $m_{\eta}-m_\chi>7~{\rm GeV}$
    (DELPHI~\cite{Abdallah:2003xe})
  \item  $m_{\eta}\geq 89~{\rm GeV}$, assuming $m_{\eta}-m_\chi>8~{\rm GeV}$
    (ALEPH~\cite{Heister:2002hp}).
\end{itemize}
Again, as long as $\mu\approx 1.1$, our toy model is not affected by these
limits; only much larger mass splittings are already partially ruled out.
\medskip

\emph{Oblique parameters.}
The new scalar particles carry charges under the electroweak group and can in
principle modify the oblique parameters $S$, $T$ and
$U$~\cite{Peskin:1990zt,Peskin:1991sw}. However, in the scenario where the
scalar particle is an SU(2)$_L$ singlet, none of the three oblique parameters
receives any exotic contribution.  \medskip

\emph{Muon $g-2$.}
The interaction which leads to the VIB process $\chi\chi\rightarrow
\mu^+\mu^-\gamma$ leads, upon closing the DM fermion loop, to a contribution
to the muon $g-2$. In the toy model presented in this work, the contribution
from DM particles to the muon $g-2$ reads (see \fex \cite{Moroi:1995yh})
\begin{align}
  \Delta a_\mu^{\rm DM}=-\frac{y^2}{16\pi^2} \frac{m^2_\mu}{m_\chi^2}F(\mu)\;,
  \label{eq:Deltag-2}
\end{align}
where $\mu\equiv(m_\eta/m_\chi)^2$ denotes the usual mass splitting, which
enters the function
\begin{align}
  F(x)=\frac{2+3x-6x^2+x^3+6x\log x}{6(x-1)^4}\;.
\end{align}
This contribution should be compared to the 3.2$\sigma$ deviation between the
Standard Model prediction and the experimental
measurement~\cite{Jegerlehner:2009ry},
\begin{align}
  a_\mu^{\rm exp}-a_\mu^{\rm SM}=(29\pm 9)\times 10^{-10}\;,
\end{align}
which is of \emph{opposite} sign.  Therefore, in our toy model the discrepancy
between the theoretical prediction of the muon $g-2$ with the experimental
measurement is larger than in the Standard Model. If we interpreted the $g-2$
anomaly as a statistical fluctuation, the total theoretical prediction should still
not deviate more than $5\sigma$ from the experimental value, which implies in
the limit $\mu\to1$ an upper bound on the coupling of $y\lesssim 1.7
(m_\chi/100\GeV)$. The corresponding upper limit on the three-body
annihilation cross-section reads
\begin{align}
  \langle \sigma v \rangle_\text{3-body} < 2.5\times10^{-26}\cm^3\s^{-1}
  \Big(\frac{m_\chi}{100\,{\rm GeV}}\Big)^2\;.
\end{align}
Comparing this to our VIB limits shown in Fig.~\ref{fig:vibBoosts}, we find
that---within our toy-model and the above assumptions---the $g-2$ constraints
are typically weaker than what we get from the spectral analysis of gamma-ray
fluxes. Alternatively, additional particles could exist that generate a positive
contribution to the muon $g-2$ which compensates the negative contribution
from the DM particles, thereby bringing the theoretical prediction closer to the
experimental value.
\medskip

\subsection{Anti-proton observations and direct WIMP searches} 
\label{sec:limits2}
In our toy model, the presence of a light colored scalar $\eta$ opens up the
gluon-VIB channel $\chi\chi\to \bar{f} f g$; in general, this process has a
$\sim100$ times larger cross-section than $\chi\chi\to\bar{f} f\gamma$,
because instead of $\alpha_\text{em}$ the strong coupling $\alpha_s$ enters
the diagrams~\cite{Drees:1993bh, Barger:2006gw, Garny:2011ii, Asano:2011ik}.
Hence, the limits on $\chi\chi\to\bar b b \gamma$ that can be obtained by
constraining the corresponding process $\chi\chi\to\bar b b g$ with cosmic-ray
anti-proton observations are quite significant for small values of $\mu$;
depending on the details of cosmic-ray propagation, they can be comparable to
or even stronger than our gamma-ray limits from the VIB
search~\cite{Garny:2011ii, Asano:2011ik}. Furthermore, $\chi\chi\to \bar{b} b
g$ contributes to the gamma-ray dwarf limits in Fig.~\ref{fig:vibBBAR}
\cite{Asano:2011ik}.

The leptophilic models discussed in this paper also give rise to an antiproton
flux through the annihilation channel $\chi\chi\to\bar f f Z$ which, when
kinematically open, has a cross-section comparable to $\chi\chi\to\bar
ff\gamma$~\cite{Kachelriess:2009zy, Bell:2011eu, Ciafaloni:2011sa,
Bell:2011if, Garny:2011cj, Ciafaloni:2011gv, Garny:2011ii}. For these models,
our gamma-ray limits are one order of magnitude stronger than the antiproton
limits in the mass range $m_\chi\sim 100$--$300\GeV$.

Lastly, for our bottom quark scenario, limits from direct DM searches may
become very relevant, as the scattering between $\chi$ and the detector nuclei via
an intermediate $\eta$ becomes resonant for small values of
$\mu$~\cite{Hisano:2011um}.  

\section{Conclusions}
\label{sec:conc}
One of the main challenges that are encountered when searching for signatures
from DM annihilation in the cosmic gamma-ray flux is the discrimination
between a possible signal and the astrophysical background. Very pronounced
spectral features, like gamma-ray lines or photons from internal
bremsstrahlung, play  a key role in this context because it would be difficult
to attribute them to astrophysical processes. Previous searches mostly
concentrated on gamma-ray lines; here, we instead analysed the effects of
internal bremsstrahlung which is  intrinsically more promising due to the
larger rates that are expected.
 
To this end, we defined a simple toy model (which can be considered as a
subset of the MSSM particle content) with the important feature to generate an
intense gamma-ray signature from virtual internal bremsstrahlung. We 
performed a dedicated search for this signature in 43 months of data from the
Fermi Large Area Telescope, building on standard methods for gamma-ray line
searches. In order to determine the optimal  target regions for our spectral
analysis, we introduced a new adaptive method that takes into account
different conventional and cuspy DM halo profiles, see
Fig.~\ref{fig:targetRegions}.  We believe that this method will turn out to be
useful even in other contexts, essentially whenever spectral features are
being looked for at the statistical limits of the detector.

For our toy model, in case of leptonic final states, we find \emph{upper
limits} on the annihilation cross-section that are stronger than what can be
obtained from dwarf galaxies or collider searches, see
Figs.~\ref{fig:vibBoosts}, \ref{fig:vibMUMU}, \ref{fig:vibTAUTAU} and
\ref{fig:vibBBAR}. Our limits are still about an order of magnitude too weak
to constrain annihilation cross-sections  naively expected for a thermal relic
(assuming a standard NFW profile for the DM density and no enhancement of the
annihilation rate due to substructures).  However, future prospects to do so
are quite good~\cite{Bringmann:2011ye}. 

We also find a weak indication for an internal bremsstrahlung \emph{signal},
see Fig.~\ref{fig:fit}, corresponding to a DM mass of $m_\chi=149\pm4\
^{+8}_{-15}\GeV$ and an annihilation rate of $\langle\sigma
v\rangle_{\chi\chi\to\bar f f \gamma}= (6.2\pm1.5\ ^{+0.9}_{-1.4})\times
10^{-27}\cm^3\s^{-1}$ (we note here that the same signal can also be fitted 
with a conventional gamma-ray line at around 130 GeV and with a 
cross-section of about $\langle \sigma v\rangle_{\chi\chi\to\gamma\gamma}
\sim10^{-27}\cm^3\s^{-1}$, and refer to Ref.~\cite{Weniger:2012tx} for a more 
detailed discussion). After taking into account the look-elsewhere effect, the 
signal significance is about $3.1\sigma$ (without, it is $4.3\sigma$). We have 
performed several statistical tests of our method and deem that a purely 
instrumental effect would be a very unlikely explanation for this signal. As 
stressed above, however, such a large radiative annihilation cross-section 
would be too large to be compatible  with naive expectations for a thermal relic, 
at least in our simple scenario and for standard cosmological  and astrophysical 
assumptions. In any case, although the observation of a VIB or line-like feature 
from DM annihilation is a fascinating possibility, we caution that more data and 
a much more refined analysis, taking into account all systematics of the LAT,
are required to reject or confirm this interpretation.

\begin{acknowledgments}
  We are very grateful to Mathias Garny, Dieter Horns and Gabrijela Zaharijas
  for useful discussions and comments. This work was initiated during the
  \emph{Dark Matter and New Physics} workshop at the Kavli Institute for
  Theoretical Physics China, Beijing; HX, AI and CW thank for warm hospitality
  and an inspiring environment. The work of AI and SV was partially supported
  by the DFG cluster of excellence \emph{Origin and Structure of the
  Universe}. SV acknowledges support from the DFG Graduiertenkolleg
  \emph{Particle Physics at the Energy Frontier of New Phenomena}. TB
  acknowledges support from the German Research Foundation (DFG) through Emmy
  Noether grant BR 3954/1-1. CW acknowledges partial support from the European
  Union FP7 ITN INVISIBLES (Marie Curie Actions, PITN-GA-2011-289442). In our
  analysis, we made use of IPython~\cite{IPython}, SciPy~\cite{SciPy},
  PyFITS~\cite{PyFits} and PyMinuit~\cite{PyMinuit}.
\end{acknowledgments}

\appendix
\section{Selection of  target regions}
\label{apx:targetRegions}
In order to define our target regions, we proceed as follows: We produce a
two-dimensional equirectangular count map that covers the region defined by
$|b|<90^\circ$ galactic latitude and $-90^\circ<\ell<90^\circ$ galactic
longitude; the pixel size is $\Delta b = \Delta\ell=1^\circ$; each pixel $i$
contains the number of gamma-ray events $c_i$ that were measured with energies
in the range 1--40 GeV (note that these $c_i$ are completely unrelated to the 
$c_j$ introduced in Section \ref{sec:specan}). Since gamma-ray fluxes drop 
rapidly with energy, the countmap produced
in this way is completely dominated by events with energies close to 1 GeV.
We will use this count map as a simple but efficient template to estimate the
spatial distribution of background events in the sky. Since for our spectral
analysis we are actually only interested in DM signatures with energies
\textit{larger} than 40 GeV, we make the assumption that the fluxes measured
at $\sim1\GeV$ resemble the spatial distribution of background events at
higher energies reasonably well (instead, one could use models for the diffuse
emission of the galaxy; this is left for future work).  For each pixel $i$, we
then calculate the number of expected signal events $\mu_i$ as predicted by
\eqref{eqn:fluxADM} in the energy range $40$--$300\GeV$. This number depends
on the adopted dark matter profile (namely the value of the inner slope
$\alpha$), and is only defined up to an overall normalization, because we
leave the annihilation cross-section and the actual annihilation spectrum
unspecified at this point.  Note that the finite angular resolution of Fermi
LAT, which is $\Delta\theta\lesssim 0.2^\circ$ above $40\GeV$, is neglected.
The signal-to-noise level in each pixel can now be estimated as
$\mathcal{R}_i\propto\mu_i/\sqrt{c_i}$ because a potential signal will likely 
only be a subdominant perturbation of the background fluxes, \ie $\mu_i\ll c_i$. 

Under the above assumptions, the optimal target region is given by the set of
pixels $\mathcal{T}_o$ for which the \emph{overall} signal-to-noise ratio
$\mathcal{R}_{\mathcal{T}_o}$, defined as
\begin{align}
  \mathcal{R}_{\mathcal{T}_o}\equiv\frac{\sum_{i\in\mathcal{T}_o} \mu_i }{
  \sqrt{\sum_{i\in\mathcal{T}_o} c_i}}\;,
\end{align}
is maximized. Finding the true $\mathcal{T}_o$ requires in principle a scan
over all $\sim2^{180^2}$ possible pixel combinations. Since this is
unfeasible, we obtain an approximate $\mathcal{T}_o$ by using the following simple
algorithm: 
\begin{enumerate}
  \item We start with an empty set $\mathcal{T}$ and include only the one
    pixel with the largest individual signal-to-noise level $\mathcal{R}_i$ as
    seed (this pixel typically lies at the galactic center).
  \item For each pixel that is not already in $\mathcal{T}$, we calculate how
    $\mathcal{R}_\mathcal{T}$ changes when this pixel is added; then all
    pixels that are found to improve $\mathcal{R}_\mathcal{T}$ are added to
    $\mathcal{T}$ at once.
  \item For each pixel in $\mathcal{T}$, we calculate how
    $\mathcal{R}_\mathcal{T}$ changes when this pixel is removed; then all
    pixels for which an increase of $\mathcal{R}_\mathcal{T}$ is found are
    removed from $\mathcal{T}$ at once.
  \item We repeat steps 2 and 3 until the number of pixels in $\mathcal{T}$
    remains constant.
  \item We remove remaining small regions in $\mathcal{T}$ that are not
    directly connected to the (always dominating) region at the galactic
    center.
\end{enumerate}
The target region obtained in this way is a very good approximation to the
optimal region, $\mathcal{T}_o\simeq\mathcal{T}$. Note that the removal of
remaining small regions in point 5 does not affect our results but merely
cleans up the borders of the derived target region, and that the final regions
are practically independent of the position of the seed pixel in point 1.  The
regions obtained in this way for the different adopted DM profiles are shown
in Fig.~\ref{fig:targetRegions}, and will be used during our spectral
analysis.

\section{Details on the statistical analysis}
\label{apx:details}

\begin{figure}[t]
  \begin{center}
    \includegraphics[width=0.8\linewidth]{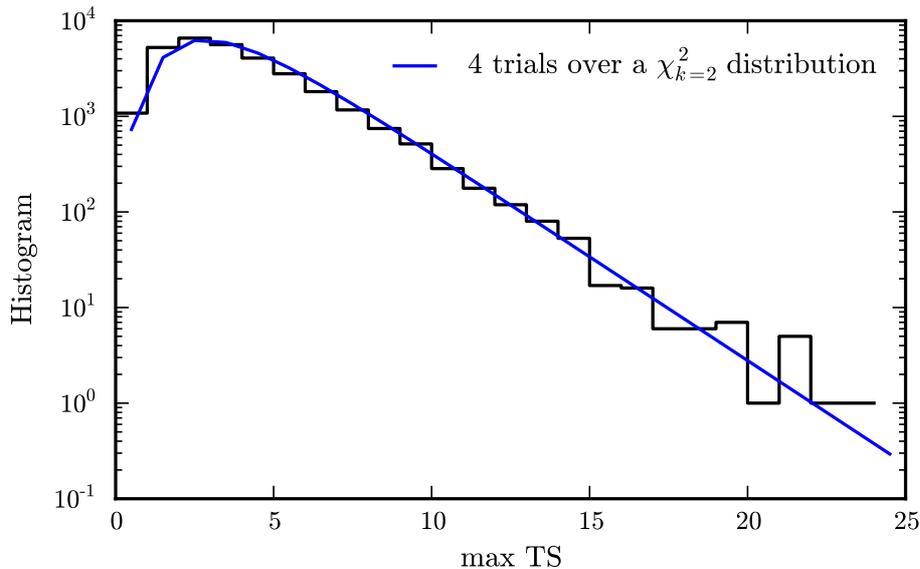}
  \end{center}
  \caption{Histogram of the maximal TS values obtained from a 
  subsampling analysis of the Fermi LAT data in the hemisphere pointing
  towards the galactic anticenter. The blue line shows the theoretical
  distribution that we used to calculate the look-elsewhere effect.}
  \label{fig:MC}
\end{figure}

\begin{figure}[t]
  \begin{center}
    \includegraphics[width=0.7\linewidth]{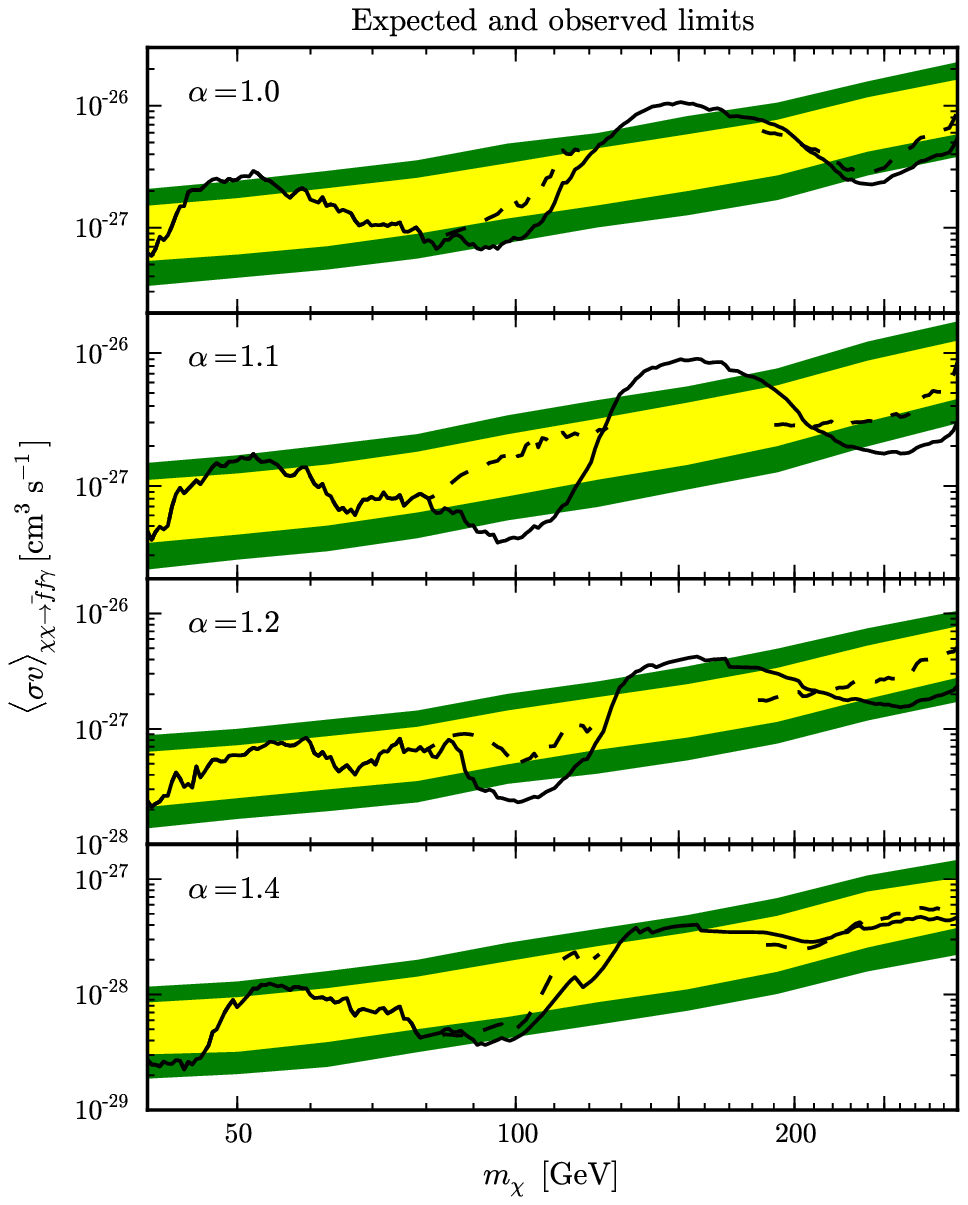}
    \caption{Experimental sensitivity for VIB-like features compared to actual
    limits. The yellow (green) bands show the expected limits at $68\%$
    ($95\%$) CL, see text for details. The black solid line shows the actually
    observed limits; these limits are significantly weaker than expected at
    dark matter masses around 150 GeV. The dashed black line shows for
    comparison the limits obtained when removing the data between 115 to 145
    GeV from our fits.}
  \label{fig:sensi}
  \end{center}
\end{figure}

In order to study the sampling distribution of the test statistic TS
\eqref{eqn:TS} in absence of a signal, we performed a \emph{subsampling analysis} 
of Fermi LAT data. To this end, we extracted the
gamma-ray events measured in the hemisphere pointing towards the anti-galactic
center, with longitudes $|\ell|>90^\circ$. Any signal from DM annihilation
should be significantly suppressed in this direction. From these events we
generate 30000 random sample spectra, with the Poisson expectation values in
each energy bin given by $\mu_j=f c_j$. Here, $c_j$ is the number of actually
measured events in bin $j$, and $f=0.13$ is adjusted such that the total number
of events above 1 GeV in each random sample is $\sim4\times 10^5$ (in Reg1 and
Reg2 the number of events are $5.8\times 10^5$ and $2.7\times10^5$,
respectively). In the limit of large event numbers, this is equivalent to
subsampling the energy distribution of the measured events with replacement.
In each of these sample spectra, we search for VIB features like discussed
above and record the largest TS value found. The histogram of the maximal TS
values that are obtained in this way is shown in Fig.~\ref{fig:MC}. There, we
also show the distribution that one obtains when selecting the maximum from 4
trials over a $\chi^2_{k=2}$ distribution. The agreement is very good, and we
used this distribution when calculating the look-elsewhere effect above.
\medskip

In Fig.~\ref{fig:sensi} we show the observed limits (black solid lines) in
comparison with the limits that are expected at $68\%$ (yellow) and $95\%$
(green) CL. We derived these expected limits from 2000 mock data samples that
were generated from the null model. In Reg1 to Reg3, the limits at
$m_\chi\simeq150\GeV$ are significantly weaker than the expectation; this
corresponds to the large TS values in the left panel of Fig.~\ref{fig:fit}.
On the other hand, the relatively strong limits at around $m_\chi\approx
100\GeV$ and $m_\chi\approx 250\GeV$ are a consequence of the $150\GeV$
excess, which influences the background fits. To illustrate this, we show by
the dashed black lines the limits that we obtain when removing all data
between 115 to 145 GeV (where the excess is most pronounced) from the fits; in
this case the limits remain in the expected range.

\bibliographystyle{JHEP}
\bibliography{}

% Throughout we use mb = 5 GeV

\end{document}